\documentclass[letterpaper,11pt]{article}
\pdfoutput=1 

\usepackage{jheppub} 

\usepackage{bm} 
\usepackage{hyperref}
\usepackage{amsmath,amssymb,amsthm} 
\usepackage{mathtools}
\usepackage{tensor} 
\usepackage{braket} 
\usepackage{multirow}
\usepackage{parskip} 
\usepackage{eufrak}

\DeclareMathOperator{\Tr}{Tr}

\newcommand{\dd}{\text{d}}

\newcommand{\mpl}{M_{\text{Pl}}}
\newcommand{\meff}{m_{\text{eff}}}
\newcommand{\cO}{c_{\mathcal{O}}}
\newcommand{\ANH}{A_{\text{NH}}}
\newcommand{\BNH}{B_{\text{NH}}}
\newcommand{\FNH}{F_{\text{NH}}}

\newcommand{\gcrit}{\gamma_{\text{crit}}}

\newcommand{\ggr}{\gamma_{\text{GR}}}
\newcommand{\geft}{\gamma_{\text{EFT}}}
\newcommand{\gmarg}{\hat{\gamma}}

\def\be{\begin{equation}}
	\def\ee{\end{equation}}
\def\ba{\begin{eqnarray}}
	\def\ea{\end{eqnarray}}
\def\d{\mathrm{d}}
\def\({\left(}
\def\){\right)}
\def\({\left(}
\def\){\right)}
\def\pp{{$pp$}}
\def\sp{{$sp$}}
\DeclareMathAlphabet{\mathpzc}{OT1}{pzc}{m}{it}
\def\ccO{\mathcal{O}}

\allowdisplaybreaks 
\hypersetup{
	colorlinks=true,
	allcolors=[rgb]{0.28, 0.24, 0.75}} 

\numberwithin{equation}{section}

\title{Deformations of Extremal Black Holes and the UV}

\author[a]{Calvin Y.-R. Chen}
\author[a,b,c]{\!\!, Claudia de Rham}
\author[a,b,c]{and Andrew J. Tolley}

\affiliation[a]{Department of Physics, Blackett Laboratory, Imperial College, London, SW7 2AZ, UK}
\affiliation[b]{Perimeter Institute for Theoretical Physics, 31 Caroline St N, Waterloo, ON, N2L 2Y5, Canada}
\affiliation[c]{CERCA, Department of Physics, Case Western Reserve University, 10900 Euclid Ave, Cleveland, OH 44106, USA}

\emailAdd{calvin.chen16@imperial.ac.uk}
\emailAdd{c.de-rham@imperial.ac.uk}
\emailAdd{a.tolley@imperial.ac.uk}

\abstract{
It has recently been noted that deformations of extremal AdS black holes in four and higher dimensions are generically non-smooth or singular on the horizon.
Further, it was found that certain deformations of asymptotically flat extremal black holes are marginal, causing the nature of the horizon to be strongly sensitive to UV corrections --- in particular, Wilson coefficients associated with consistent UV completions generically lead to worse behaviour on the horizon.
In this work, we extend the discussion of deformed horizons in the presence of UV corrections to extremal charged black holes in AdS, where we find a tower of marginal perturbations for different black hole masses.
We argue that the apparent UV sensitivity of marginal modes is, in fact, a feature of the UV theory which is correctly reproduced by the EFT, and illustrate this with explicit UV completions confirming the validity of the EFT.
We demonstrate that the same holds for a scalar-Maxwell EFT with known UV completion.
In the gravitational case, the sign of EFT corrections to marginal perturbations is generally connected with the signs implied by positivity bounds, with UV completions generically leading to worse behaviour on the horizon. We conjecture that this result is more generic, and use this to derive more general positivity bounds motivated by the weak gravity conjecture, which we illustrate with further evidence.
}

\begin{document}

\maketitle

\section{Introduction \label{sec: intro}}

Black holes (BHs) are one of the best studied objects in general relativity (GR), that have provided the impetus leading to a number of breakthroughs in our understanding of gravity.
Particularly striking is the prediction of its own breakdown through the generic formation of curvature singularities \cite{Penrose:1964wq}.
In practical terms, this means that in order to understand dynamics at very high curvature (sometimes qualified as high energies or small length scales), a UV completion to GR, \textit{i.e.} a theory of quantum gravity, is necessary.
An example of one such UV completion is string theory --- it is a UV finite theory that contains GR in its low-energy effective field theory (EFT) and predicts specific higher-order corrections to it \cite{Metsaev:1986yb}.
More generically, we can parameterise UV corrections to the low-energy description with generic higher-order corrections to the Einstein-Hilbert action \cite{Wilson:1973jj,Donoghue:1994dn,Donoghue:1995cz,Burgess:2003jk,Donoghue:2012zc,Weinberg:2018apv}.

In more recent years, it was realised that EFTs also need to satisfy constraints beyond symmetry and generic low-energy consistency in order to have any chance to be embeddable in a consistent UV completion, see \cite{deRham:2022hpx} for a review.
For non-gravitational theories, it is now well-understood that properties such as unitarity, locality, Lorentz invariance, and causality are non-trivially retained under RG flow and put constraints on possible low-energy physics \cite{Pham:1985cr,Ananthanarayan:1994hf}.
For example, causality leads to positivity constraints on derivatives of the scattering amplitude through analyticity of the $S$-matrix \cite{Adams:2006sv,Nicolis:2009qm,Giddings:2009gj,Bellazzini:2014waa,Baumann:2015nta,Bellazzini:2015cra,
Cheung:2016wjt,Bellazzini:2016xrt,deRham:2017avq,deRham:2017zjm,deRham:2018qqo,Bonifacio:2018vzv,Alberte:2019xfh,
Tolley:2020gtv,Tokuda:2020mlf,Arkani-Hamed:2020blm,Caron-Huot:2020cmc,Sinha:2020win,Bellazzini:2020cot,
Herrero-Valea:2020wxz,Noumi:2021uuv,Du:2021byy,Chiang:2021ziz,Haldar:2021rri,Raman:2021pkf,Chowdhury:2021ynh,Chowdhury:2022obp,
Bern:2021ppb,Herrero-Valea:2022lfd,Melville:2022ykg,Noumi:2022zht,Chiang:2022jep},
and directly manifests itself within the low-energy theory through the absence of superluminal propagation in the form of a resolvable and negative time delay \cite{Nussenzveig:1972tcd,Gao:2000ga,Hollowood:2007ku,Shore:2007um,Camanho:2014apa,Hartman:2015lfa,Camanho:2016opx,Goon:2016une, Hinterbichler:2017qcl,AccettulliHuber:2020oou,Serra:2022pzl,CarrilloGonzalez:2022fwg,CarrilloGonzalez:2023cbf,Cremonini:2023epg,Hong:2023zgm,Chen:2023rar}.
This is not straightforward for gravitational theories, with one of the key realisations being that known flat-space bounds are at the very least weakened in the presence of dynamical gravity \cite{Bjerrum-Bohr:2014lea,Haring:2022cyf,Alberte:2020jsk,Alberte:2020bdz,Henriksson:2022oeu,deRham:2019ctd,deRham:2020zyh,Alberte:2021dnj,Caron-Huot:2021rmr,Caron-Huot:2022jli,deRham:2022gfe, Caron-Huot:2022ugt,Aoki:2023khq,Hamada:2023cyt,Caron-Huot:2024tsk}.
Related to this, the so-called ``swampland program" aims to delineate the boundary between the string landscape and eponymous swampland --- in other words, find qualitative restrictions directly on quantum gravity --- informed by explicit constructions in string theory \cite{Vafa:2005ui,Saraswat:2016eaz,Palti:2019pca,Agmon:2022thq}.
Two of the strongest and most well-known of these conjectures are the ``no global symmetries" and ``weak gravity" conjectures (WGC) --- these were first discussed in \cite{Banks:1988yz} and \cite{Arkani-Hamed:2006emk} respectively, but have regained significant interest in recent years \cite{Cheung:2014ega, Cheung:2018cwt,Aalsma:2020duv,Hamada:2018dde,Alberte:2020jsk,Alberte:2020bdz,Arkani-Hamed:2021ajd,Alim:2021vhs, Cota:2022maf, Harlow:2022ich,Henriksson:2022oeu,Rudelius:2022gyu, deBrito:2023myf,Alipour:2023yiz}.
Qualitative arguments for both of these conjectures crucially involve the existence of extremal BHs and their ability to decay without becoming \mbox{(super-)extremal}.
Extremal BHs and black branes have also played a prominent role as non-perturbative states in superstring theory for the microstate counting of BH entropy, solitonic solutions preserving partial supersymmetry in supergravity, or AdS/CFT \cite{Strominger:1996sh,Maldacena:1996ky, Maldacena:1997re}.

The surprising feature of extremal BHs that we are concerned with lies in (or rather on) their horizons.
Horizons of Schwarzschild BHs in $D \geq 4$, and in fact due to uniqueness all asymptotically flat and stationary BHs in $D=4$, have smooth horizons.
However, the expectation that this persists when any of these conditions is relaxed, \textit{i.e.} for solutions in higher dimensions or less symmetry, turns out to be wrong \cite{Welch:1995dh, Kimura:2014uaa}.
The simplest and most famous counterexample to this is the Majumdar-Papapetrou
solution, describing multiple extremal BHs, which has horizons that have finite regularity class \cite{Candlish:2007fh, Gowdigere:2014aca, Gowdigere:2014cqa}.

Recently, by studying particular deformations of extremal Anti-de Sitter (AdS) Reissner-Nordstr\"om (RN) BHs, \cite{Horowitz:2022mly, Horowitz:2022leb,Cano:2024bhh} established that this feature is generic to ``almost all" AdS BHs.
These deformations are stationary metric perturbations $h$ that exhibit critical scaling of the form
\begin{equation}
	h \sim \rho^{\gamma},
	\label{eq: critical scaling}
\end{equation}
in an orthonormal frame with the horizon located at $\rho=0$. Physically these are tidal deformations induced by matter sources external to the BH, and an explicit nonlinear example are provided by the Majumdar-Papapetrou multi-black holes as outlined in appendix~\ref{subapp: majumdar-papapetrou}.
For $0<\gamma \neq 1<2$, all scalar curvature invariants are finite but tidal forces diverge.
If $\gamma<0$ curvature invariants diverge, although in practice we can only conclude the breakdown of linear perturbation theory. It is possible that a stable, less symmetric near-horizon geometry emerges non-linearly , although this was explored in \cite{Horowitz:2022leb} and no such stable solutions has been found. There is however no reason one should necessarily expect the evolution to asymptote to a static endpoint.

Further, \cite{Horowitz:2023xyl} identified a marginal deformation of asymptotically flat RN BHs in $D=5$ which opens up the possibility for UV sensitivity: The GR contribution to $\gamma$ vanishes and, as diagnosed in \cite{Horowitz:2023xyl,Horowitz:2024dch}, since the EFT corrections (which are governed by UV physics) determine the sign of $\gamma$ and therefore the presence of singularities on the horizon, these systems may appear to indicate a level of UV sensitivity.
Superficially this seems to undermine the spirit of EFTs, namely that the low-energy physics should be largely insensitive to the precise details of the UV physics.
Perhaps most interestingly, the higher-derivative operators arising from Wilson coefficients satisfying the expectations from the WGC lead to negative corrections to $\gamma$.

In this work, we consider aforementioned deformations to the near-horizon region of extremal charged BHs, and extend the analysis of deformed horizons in the presence of UV corrections to AdS RN BHs and systematically study the effect of UV physics on marginal deformations thereof.

Before proceeding, let us first describe the set-up in more detail and illustrate the argument with a concrete example.
In a theory of gravity which includes GR and higher-derivative EFT corrections, the effect of said corrections on perturbations in the vicinity of extremal horizons resums into the exponent of \eqref{eq: critical scaling}.
These take the form of
\begin{equation}
	\gamma = \ggr + \geft,
	\label{eq: critical exponent resummed}
\end{equation}
such that $\gamma$ reduces to $\ggr$, by which we mean the $\gamma$ implied by the Einstein-Maxwell theory, in the absence of EFT or higher-derivative corrections.
Taken at face value, curvature invariants blow up on the horizon when $\gamma<0$, but more precisely perturbation theory breaks down. By contrast if $\gamma>0$ perturbation theory remains under control and the singularity is mild. So $\gcrit:=0$ constitutes a marginal case or critical value for singular behaviour\footnote{For a discussion on the distinction between $\gamma<0$ and $\gamma<2$, see section \ref{subsubsec: singularities}.}.
When the GR contribution is marginal, \textit{i.e.} for $\ggr = 0$, the EFT contribution seems to dominate if non-zero, and the system appears to be UV sensitive, in effect amplifying the UV physics \cite{Horowitz:2023xyl}.
It turns out that for extremal AdS RN BHs, there exists an infinite number of deformations to the near-horizon geometry for which this is exactly the case.

\subsection{Sign dependence}

As a specific example to start with, we may consider Einstein-Maxwell theory with zero cosmological constant in $D=5$, described by the following EFT-corrected action
\begin{equation}
	S_{\text{IR}} = \int \dd^{5}x \sqrt{-g} \left[\frac{1}{2\kappa}R - \frac{1}{4}F_{\mu\nu}F^{\mu\nu} + c\frac{\kappa}{\Lambda^{2}} \left(F_{\mu\nu}F^{\mu\nu}\right)^{2}\right] ,
	\label{eq: example ir action}
\end{equation}
with $\kappa=\mpl^{-3}$ and cutoff $\Lambda$.
From the EFT perspective, this theory is of course highly tuned since we anticipate additional non-zero operators, but the specific EFT correction itself is generic and arises for example at loop level from integrating out massive charged states, and at tree level from non-minimally coupled chargeless scalars.
Perturbatively in the Wilson coefficient $c$, the theory admits asymptotically flat, static, and spherically symmetric background solutions with $\mathrm{AdS}_{2} \times S^{3}$ near-horizon geometry in the extremal limit.
On top of these, perturbations (parameterized by their respective multipole $\ell$) take the form of scaling solutions in \eqref{eq: critical scaling} with exponent \eqref{eq: critical exponent resummed}, where
\begin{equation}
		\ggr = \frac{1}{2}\left(-1 + \left|\ell - 1\right|\right), \quad \geft = -\frac{c}{\Lambda^{2}r_{H}^{2}}\frac{6\ell(\ell+2)}{\ell-1}\,,
\end{equation}
and $r_{H}$ is the extremal BH horizon radius.
As noted in \cite{Horowitz:2023xyl}, the quadrupole ($\ell=2$) is a marginal deformation for which
\begin{equation}
	\gmarg := \gamma\big|_{\ggr=0} = -48\frac{c}{\Lambda^{2}r_{H}^{2}}.
	\label{eq: example ir marginal scaling}
\end{equation}
This is negative when $c> 0$, which is the sign expected from non-gravitational positivity bounds and the WGC, and \textit{vice versa}.
More precisely, in the decoupling limit $\mpl \rightarrow \infty$, $\kappa \rightarrow 0$, we would still expect finite $ \left(F_{\mu\nu}F^{\mu\nu}\right)^{2}$ corrections coming from loops with charged massive states.
If the lowest such state has mass $\Lambda$ and charge $q$ then we expect $c \kappa \approx q^2 \Lambda^{-2}$.
Furthermore, Minkowski spacetime positivity bounds coming from photon scattering amplitudes would imply
\begin{equation} \label{dbound}
c >0 \,
\end{equation}
in that decoupling limit.
In other words, within our truncated theory, we find that ``consistent" low-energy physics is associated with a definite sign for the correction $\gmarg$.

When applied to gravitationally coupled theories, positivity bounds are weakened to allow a small (Planck scale suppressed) amount of negativity \cite{Alberte:2020jsk,Alberte:2020bdz,Henriksson:2022oeu,deRham:2019ctd,deRham:2020zyh,Alberte:2021dnj,Caron-Huot:2021rmr,Caron-Huot:2022jli,deRham:2022gfe, Caron-Huot:2022ugt,Aoki:2023khq,Hamada:2023cyt,Caron-Huot:2024tsk},
so we expect in $D$ dimensions Eq.~\eqref{dbound} to be replaced by,
\be
c>-|{\cal O}(\kappa \Lambda^{{D-2}})|\, .
\ee
Thus a more refined estimate of the bound is
\be
\gmarg <+  \frac{|{\cal O}(\kappa \Lambda^{{D-2}})|}{\Lambda^{2}r_{H}^{2}} .
\ee

\subsection{Apparent UV sensitivity}
As discussed previously, some curvature invariants blow up on the horizon when $\gamma<0$, so
when the GR contribution is marginal, \textit{i.e.} for $\ggr = 0$, the sign of the coefficient $\hat \gamma$ is set by the EFT corrections.
It is tempting to view this apparent sensitivity to UV physics as signalling a \textit{breakdown of the EFT}.

\paragraph{Validity of the EFT:} The validity of the EFT requires multiple checks.
\begin{itemize}
\item A first check on the validity of the EFT is to show that higher order corrections to $\gamma$ in the EFT expansion, namely those of order $(r_H\Lambda)^{-4}$ are small as compared to the leading ones.
This is intrinsically a check on the validity of the EFT at the background level.
    With the na\"ive expectation that the dimensionless Wilson coefficients $c$ are of order unity, this will be true provided $r_H \gg \Lambda^{-1}$, which is required to trust the background solution.
We shall construct specific (partial) UV completions later that demonstrate this in explicit examples. However, by itself this statement is insufficient.
\item
In addition, validity of the EFT should also be accounted for at the perturbed level, and we also need to check that the perturbations do not exit the regime of validity.
This occurs when curvature invariants constructed within the deformed geometry, including the perturbations, become large.
For $\gamma<0$, this inevitably occurs before we reach the extremal horizon.
For an EFT with a Lorentz-invariant energy cut-off $\Lambda$ on an extremal BH background with horizon radius $r_{H}$, we show that this breakdown occurs when the metric perturbations reach
\begin{equation}
	h \sim h_{0} \rho^{\gamma} \sim (\Lambda r_{H})^{2}\,,
\end{equation}
in some orthonormal frame.
As already noted, for the EFT to be valid even at the level of the background, we also require that $\Lambda r_{H} \gg 1$.
\end{itemize}

\paragraph{Validity of perturbation theory:} On the other hand, metric perturbation theory itself breaks down when the quadratic action in perturbations no longer dominates over higher-order terms in the action.
We confirm that this happens when
\begin{equation}
	h \sim h_{0} \rho^{\gamma} \sim 1\,,
\end{equation}
as expected.

Evidently, for $\gamma<0$ metric perturbation theory breaks down first\footnote{Whenever $\Lambda r_H\lesssim \mathcal{O}(1)$, the BHs are so small that they lie outside the regime of validity of the EFT in the first place irrespectively of how charged and how close to extreme they are.} whenever $\Lambda r_H\gg 1$.
In other words, we first lose control over metric perturbation theory, long before the EFT is out of control and we reach the horizon.
In the related $D=5$ example considered in \cite{Horowitz:2022leb}, where $\gamma<0$ not via EFT corrections but due to the asymptotic AdS boundary conditions, the breakdown of perturbation theory is shown to signal an instability of the IR fixed point and suggests that RG flow leads to a different IR theory likely with less symmetry.

\subsection{(Partial) UV completion}
To make the previous point more concrete, we may consider the following action as an illustrative example
\begin{equation}
	S_{\text{UV}} = \int \dd^{5}x \sqrt{-g} \left[\frac{1}{2\kappa}R - \frac{1}{4}e^{\alpha\phi}F_{\mu\nu}F^{\mu\nu} -\frac{1}{2 \kappa}(\nabla\phi)^{2}-\frac{1}{2\kappa}m^{2}\phi^{2}\right]\,,
	\label{eq: example uv action}
\end{equation}
describing Einstein-Maxwell theory coupled to a (non-canonically normalized) massive dilaton $\phi$ with mass $m$ in $D=5$.
It is a partial UV completion of the effective action \eqref{eq: example ir action} --- more precisely, when the dilaton is heavy relative to the background curvature, \textit{i.e.} $m^{2}r_{H}^{2} \gg 1$, this is the tree-level effective action we obtain from integrating out said massive dilaton in \eqref{eq: example uv action}, with $c=\alpha^2/32$ and $\Lambda= m$.

However, there is no reason we have to think of this as an EFT correction as the theory is perfectly well-behaved in the UV --- it is described by a two-derivative action which at least in form universally arises as the bosonic sector in string theory and its supergravity \cite{Deligne:1999qp, Freedman:2012zz, Blumenhagen:2013fgp}.
Perturbatively in $\alpha$ this also admits $\mathrm{AdS}_{2} \times S^{3}$ Robinson-Bertotti background solutions \cite{Robinson:1959ev, Bertotti:1959pf}, on top of which perturbations take precisely the form of the scaling solutions in \eqref{eq: critical scaling} with scaling exponent
\begin{equation}
	\gamma = \frac{1}{2}\left(-1 + \left|\ell - 1\right|\right) - \frac{\alpha^{2}}{32m^{2}r_{H}^{2}} \frac{6\ell(\ell+2)}{\ell-1} \frac{1-\frac{4\ell(3\ell-5)}{(\ell+1)m^{2}r_{H}^{2}}}{1+\frac{4\ell}{m^{2}r_{H}^{2}}}.
\end{equation}
The leading term is what we would obtain as the uncorrected $\ggr$ in pure Einstein-Maxwell theory, so the $\mathcal{O}(\alpha^{2})$-term is a correction to this.
Let us make the trivial observation that, depending on the parameters of the theory $(\alpha,m)$ and the solution $(\ell, r_{H})$, the exponent $\gamma$ can take either sign.
As discussed, $\gamma<0$ leads to metric perturbations blowing up at the horizon, rendering metric perturbation theory out of control and possible signalling that the near-horizon will be replaced by a solution with less symmetry.

The leading-order contribution itself vanishes when $\ell = 2$, in which case the marginal scaling takes the form
\begin{equation}
	\gmarg = -\frac{3\alpha^{2}}{2m^{2}r_{H}^{2}}\frac{1-\frac{8}{3m^{2}r_{H}^{2}}}{1+\frac{8}{m^{2}r_{H}^{2}}}.
\end{equation}
As expected, this agrees with \eqref{eq: example ir marginal scaling} in the limit where $m r_{H} \rightarrow \infty$ and with appropriate identification of $c$ and $\Lambda$.
Having $\gmarg$ positive, requires $m^{2}r_{H}^{2}<8/3$ which means we cannot regard the massive dilaton as heavy, which means we have changed the low-energy theory to Einstein-Maxwell-Dilaton.
By contrast, when $\gmarg$ is negative, which occurs when $m^{2}r_{H}^{2} \gg 1$ for which the Einstein-Maxwell low-energy theory is meaningful, we see that metric perturbation theory already breaks down in the UV, and the EFT in fact captures this effect perfectly. If it was the EFT itself that had broken down, we would have expected that details of the UV completion would be sufficient to cure this.
Since the singularity persists even in the UV, validity of the EFT cannot be at blame.

For deformations of asymptotically flat RN BHs, the combination of Wilson coefficients for four-derivative EFT corrections entering the asymptotically flat extremal charge-to-mass ratio also enters the scaling exponent $\gmarg$ in the marginal case.
Therefore, Wilson coefficients satisfying the asymptotically flat WGC lead to worse behaviour on the horizons of marginal deformations to asymptotically flat extremal RN BHs.
There is an infinite tower of marginal deformations of extremal RN BHs in AdS, with different combinations of Wilson coefficients entering the marginal scaling exponent $\gmarg$.
On a speculative basis, we conjecture in Section~\ref{sec: uv sensitivity} that EFTs with consistent UV completions also lead to worse behaviour on the horizons of these marginally deformed AdS RN geometries.
We explore the implications of this so-called near-horizon negativity, along the lines of other criteria and bounds within the EFT consistency program \cite{Caron-Huot:2022ugt, CarrilloGonzalez:2022fwg, CarrilloGonzalez:2023cbf, Melville:2024zjq}.

\subsection{Organisation}

The rest of the paper is organised as follows.
We start in section \ref{sec: background} by reviewing the near-horizon deformations of extremal BHs.
A scalar toy model for the deformations of interest is presented in \ref{app: scalar toy model}.
In section \ref{sec: extremal rn deformations} we take a closer look at deformations of AdS RN BHs in the presence of UV corrections, expanding on the examples from the introduction.
We then study the breakdown of EFT and metric perturbations near the deformed horizons of extremal BHs in section \ref{sec: breakdowns}.
Finally, based on a previously noted connection to the WGC, we investigate the consequences of a speculative conjecture on four-derivative corrections in Einstein-Maxwell theory in section \ref{sec: uv sensitivity}. We end with a summary and potential implications in Section~\ref{sec: conclusion}.

Details for the decompositions with respect to the sphere and $\mathrm{AdS}_{2}$ are given in appendices \ref{app: sphere decomposition} and \ref{app: nh decomposition} respectively.
appendix~\ref{app: scalar} includes the details related to the EFT of a massive scalar field non-minimally coupled to Maxwell.
Boundary conditions necessary to source the deformations to BHs are discussed in \ref{app: bcs} and the Majumdar-Papapetrou multi-BH exact solution is used as an explicit example illustrating non-linearly how physical sources lead to tidal deformations.
The scaling exponents for the deformations in GR and its extensions are given in appendix \ref{app: xrn critical exponents}.
Explicit expressions for the Einstein-Maxwell action expanded up to quartic order in perturbations are given in appendix \ref{app: em action expansion}.

\subsection{Conventions}

We work in units where $\hbar=c=1$, and on $D=d+2$-dimensional spacetime manifolds in mostly-plus signature $(-,+,\dots,+)$.
Tensors of the full $D$-dimensional manifold are indexed with letters from the Greek alphabet $\{\alpha,\beta,\dots\}$.
The Laplace-Beltrami operator is written as $\Box = g^{\alpha \beta}\nabla_{\alpha}\nabla_{\beta}$.
Tensors on the $d$-dimensional transverse space are indexed with letters from the middle of the Roman alphabet $\{i,j,\dots\}$, while tensors on the 2-dimensional orbit space are indexed with letters from the beginning of the Roman alphabet $\{a,b,\dots\}$.

We will use tildes to denote the extremal limit of a quantity and hats on coefficients to denote marginal cases, \textit{i.e.} for which the GR contribution vanishes.

For later convenience, we shall distinguish between different classes of singularities
\begin{itemize}
\item {\sp} indicate {\bf scalar polynomial} singularities,
\item {\pp} indicate {\bf parallel-propagated}  singularities.
\end{itemize}
The difference between these two classes of singularities is discussed in more detail in Section~\ref{subsubsec: singularities}.
Essentially, in this work we take the point of view that \pp--singularities lead at most to a breakdown of the worldline approach and are not necessarily signaling an apparent breakdown of the EFT.
On the other hand, \sp--singularities are potentially more problematic, but also come hand in hand with a divergence in metric perturbations (in any frame).
This means that it is essential to understand the non-linear dynamics of the system before drawing any conclusion on the relevance of these classes of singularities.

\section{Deformations of Extremal Black Holes \label{sec: background}}

In this section, we revisit deformations of extremal charged BHs from \cite{Horowitz:2022mly}.
We introduce the set-up by reviewing background solutions and the decoupling of perturbations thereof in pure Einstein-Maxwell theory.
The solutions of interest are stationary deformations to the near-horizon regions of extremal BHs, so we take particular care to discuss the various subtleties and tensions of the respective limits.
We then solve for these deformations explicitly and discuss some of their properties.

\subsection{Reissner-Nordstr\"om Black Holes: Background and Perturbations \label{subsec: rn bhs}}

Consider Einstein-Maxwell theory in $D=d+2$ dimensions with cosmological constant $\Lambda_{\rm c.c.}$, described by the following action
\begin{equation}
	S_{\text{EM}} = \int \dd^{D}x \sqrt{-g} \left[ \frac{1}{2\kappa}\left(R-2\Lambda_{\rm c.c.}\right)-\frac{1}{4}F_{\mu\nu}F^{\mu\nu}\right].
\end{equation}
In terms of the Planck mass $\mpl$, $1/\kappa = \mpl^{d}$.
For convenience, we will concentrate on asymptotically AdS spacetimes and write
\begin{equation}
	\Lambda_{\rm c.c.} = -\frac{(D-1)(D-2)}{2L^{2}},
\end{equation}
where $L$ denotes the AdS radius.
The asymptotically flat and $\mathrm{dS}$ cases can be reached via the limit $L \rightarrow \infty$ or analytic continuation $L^{2} \mapsto -L^{2}$ respectively.
In organising the EFT expansion it can be useful to work with a non-canonically normalised gauge field \mbox{$\hat A _{\mu}=\kappa^{-1/2} A_{\mu}$} so that
 \begin{equation}
	\kappa S_{\text{EM}} = \int \dd^{D}x \sqrt{-g} \left[ \frac{1}{2}\left(R-2\Lambda_{\rm c.c.}\)-\frac{1}{4}\widehat F_{\mu\nu}\widehat F^{\mu\nu} \right]\,,
	\label{eq: em action}
\end{equation}
such that the Planck scale factors out of the tree-level action.
With this normalisation, assuming $\Lambda$ is the cut-off of the EFT, loop corrections will be suppressed by powers of
\begin{equation}
 \kappa \Lambda^{D-2}=\left(\frac{\Lambda}{\mpl}\right)^{D-2} \, .
\end{equation}
The general structure of the EFT is then schematically
\be
\label{EFTexpansion}
\kappa S_{\rm EFT} = \int \d^D x \sum_{lmnp}  c_{lmnp} \(\kappa \Lambda^{D-2} \)^l  \,\( \frac{\text{Riemann}}{\Lambda^2}\)^n \( \frac{\nabla}{\Lambda} \)^{2m} \(\frac{\widehat F}{\Lambda}\)^{2 p} \, .
\ee
More explicitly, we mean to include every scalar operator (up to field redefinitions) constructed out of appropriate powers of the Riemann tensor, the field strength $\widehat F_{ab}$, and covariant derivatives, and $l$ indicates the loop order at which a particular term is expected to arise.

Returning to the leading two-derivative action \eqref{eq: em action}, the background equations of motion associated to this are
\begin{equation}
	G_{\mu\nu} + \Lambda_{\rm c.c.} g_{\mu\nu} = \kappa T_{\mu\nu}, \quad T_{\mu\nu} = F_{\mu\alpha}F\indices{_{\nu}^{\alpha}} - \frac{1}{4} F_{\alpha\beta}F^{\alpha\beta}g_{\mu\nu} \, ,
	\label{eq: einstein eq}
\end{equation}
and
\begin{equation}
	\nabla^{\mu}F_{\mu\nu} = 0.
	\label{eq: maxwell eq}
\end{equation}

For convenience, we restrict ourselves to static and spherically symmetric, and electrically charged background solutions, which correspond to the following Ansatz
\begin{subequations}
	\label{eq: sss ansatz}
	\begin{align}
		&\bar{g}_{\mu\nu}\dd x^{\mu}\dd x^{\nu} = -A(r) \dd t^{2} + \frac{\dd r^{2}}{B(r)} + r^{2} \dd \Omega_{d}^{2}, \\
		&\bar{F} = \Psi'(r) \dd t \wedge \dd r,
	\end{align}
\end{subequations}
where $\dd\Omega_{d}^{2}$ is the round metric on $S^{d}$.
The solution to the Einstein-Maxwell equations \eqref{eq: einstein eq} and \eqref{eq: maxwell eq} of this form is the AdS RN solution, and it is given by $A(r)=B(r)=f(r)$, for
\begin{subequations}
	\label{eq: ads rn bg}
	\begin{gather}
		f(r) := 1 + \frac{r^{2}}{L^{2}} - \frac{2M}{r^{D-3}} + \frac{Q^{2}}{r^{2(D-3)}}, \\
		\Psi(r) = \frac{q}{r^{D-3}},\quad \quad Q^{2}=\frac{(D-3)q^{2}\kappa}{D-2}.
	\end{gather}
\end{subequations}
To study perturbations on top of the $D$-dimensional AdS RN backgrounds, we consider small perturbations $ h:=g-\bar{g}$ and $\delta F:=F-\bar{F}$ around the background solutions and linearise the equations of motion \eqref{eq: einstein eq} and \eqref{eq: maxwell eq} in these.
Following \cite{Kodama:2003kk}, it is natural to decompose perturbations with respect to the background spherical symmetry in terms of \textit{tensor, vector, and scalar harmonics}.
Particular linear combinations of the gauge-invariant variables known as ``master variables", which mix the gravitational and electromagnetic (EM) perturbations, obey Schr\"odinger-like ``master equations"
\begin{equation}
	\Box_{2}\Phi_{M} - U_{M}\Phi_{M} = 0,\quad \Box_{2}\cdot := -\frac{1}{f} \partial_{t}^{2}\cdot + \partial_{r}(f\partial_{r}\cdot),
	\label{eq: master eq}
\end{equation}
where $M \in \{T, V_+, V_-, S_+, S_-\}$ labels the tensor, vector, and scalar modes respectively --- $V_\pm$ and $S_\pm$ are mixtures of the gravitational and gauge vector and scalar modes respectively.
Further, $U_{M}(r)$ is a mode-dependent effective potential containing the angular dependence, although the details of the effective potential are irrelevant in the upcoming discussions and we will drop the subscript $M$ for now.
See appendix \ref{app: sphere decomposition} for more details.

We shall be interested in the near-horizon region, where Schwarzschild-like coordinates are inconvenient. Instead we employ ingoing coordinates $(v,r,x^{i})$ with
\begin{equation}
	v := t + r_{*},\quad \dd r_{*} := \frac{\dd r}{f}.
\end{equation}
The perturbations now obey the master equation \eqref{eq: master eq} with
\begin{equation}
	\Box_{2}\cdot = 2\partial_{v}\partial_{r} \cdot + \partial_{r}(f \partial_{r}\cdot).
\end{equation}
Due to staticity of the background, we can use the Killing vector $k=\partial/\partial v$ to decompose the perturbations in terms of its eigenfunctions $\Phi_{M}(t,r) = e^{-i\omega v}\phi_{M}(r)$, so that \eqref{eq: master eq} reduces to the following master equation
\begin{equation}
	f \phi_{M}'' + (f'- i\omega) \phi_{M}' - U_{M} \phi_{M} = 0\,,
	\label{eq: master ode}
\end{equation}
on which we focus for the rest of this work.

\subsection{The Extremal Limit \label{subsec: extremal limit}}

AdS RN has two real horizons $r_{\pm}$ which are implicitly defined by $f(r_{\pm})=0$, while the degenerate extremal horizon $r_{H}$ is implicitly defined by $f(r_{H})=0$ and $f'(r_{H})=0$ simultaneously.
For asymptotically flat RN BHs in $D$ dimensions, this is explicitly given by
\begin{equation}
	r_{\pm}^{D-3} = M \pm \sqrt{M^2-Q^{2}},
	\label{eq: horizons}
\end{equation}
so the extremal limit is reached when
\begin{equation}
	|Q| = M \, .
	\label{eq: extremal}
\end{equation}
It will often be convenient to repackage dependence on the AdS length scale $L$ into the quantity $\sigma$
\begin{equation}
	\sigma = 1 + \frac{(D-2)(D-1)}{(D-3)^{2}} \varsigma, \quad {\rm with}\quad \varsigma = \frac{r_{H}^{2}}{L^{2}}.
\label{eq:sigma}
\end{equation}
Then, for the extremal solution, the near-horizon $f(r)$ takes the form
\be
f(r) = \frac{(D-3)^2 \sigma}{r_H^2} (r-r_H)^2+ \dots \, .
\ee

\subsection{Stationary black hole at its extremes}

We are interested in the deformations considered in \cite{Horowitz:2022mly, Horowitz:2022leb, Horowitz:2023xyl, Horowitz:2024dch} --- these are \textit{stationary} perturbations of \textit{extremal} BHs within the \textit{near-horizon} limit.
The extremal limit is in tension with both the stationary limit and the near-horizon expansion, and the order in which these expansions are performed is subtle.

\textbf{Near-horizon:} In what follows we shall solely be interested in the near-horizon limit and operate under the premise that additional matter sources/boundary conditions can, in principle, easily be considered to generate the tidal deformations of interest while maintaining acceptable $r\to \infty$ behaviour.
Our main concern is hence ensuring consistency at the horizon in the stationary and extremal limits.\\

By \textbf{stationary} perturbations, we mean solutions to \eqref{eq: master ode} with $\omega = 0$.
As an ODE, the master equation \eqref{eq: master ode} has singular points at the horizon(s) $r=r_{\pm}$ (or $r=r_{H}$ in the extremal limit), and the classification into regular and irregular singular points depends on whether the stationary and/or extremal limits are taken.
Note that stationary perturbations are a very tuned subclass (of zero measure) within the class of general perturbations. 
In fact, the zero-frequency limit of dynamical perturbations behave very differently from the actual static modes.
However, as we will see in section \ref{subsec: eft breakdown}, it turns out that non-static modes are not within the regime of validity of the EFT for any finite cut-off, \textit{i.e.} even just in pure GR, at the horizon for extremal solutions. 
%
%

Physically, \textbf{extremal} BHs should be seen as the limit of sub-extremal BHs physics.
More precisely, the near-horizon dynamics of \textit{extremal} BH perturbations are to be understood as the extremal limit of the near-horizon dynamics of \textit{non-extremal} BH perturbations.
For singular geometries, one motivation for this is the criterion established in \cite{Gubser:2000nd}.
At its core, this states that zero-temperature singular solutions are admissible if they can be reached as the zero-temperature limit of non-singular finite-temperature solutions.
The physical motivation for this is that for such geometries, the would-be naked singularities at zero-temperature would be shielded by the horizon of a regular geometry in the finite-temperature case.

\subsubsection{The Near-Horizon Limit}

In the sub-extremal case, it is convenient to parameterise the departure from extremality with the positive dimensionless coefficient $\varepsilon$
\be
\varepsilon=\frac{r_+-r_-}{r_++r_-} \, ,
\ee
and the distance to the horizon with the variable $\rho$ defined as $\rho = r-r_{+}$. %
The near-horizon and near-extremal limits then correspond to expansions in $\varepsilon$ and $\rho/r_{+}$ respectively and we shall make use of the expansion
\begin{equation}
	f(r) = f'(r_{+})\rho + \frac{1}{2}f''(r_{+})\rho^{2} + \dots
\end{equation}
within the near-horizon limit.
Notably $f'(r_{+}) = \mathcal{O}(\varepsilon)$ and vanishes in the extremal limit, so let us define $\beta=\frac{2f'(r_{+})}{\varepsilon f''(r_{+})}$ which is finite in the extremal limit.
In the near-horizon region, the master equation \eqref{eq: master ode} can be approximated by\footnote{The expansion is performed in a way to carefully account for any potential divergence and orders of limits.}
\begin{equation}
	\left(\rho^{2} + \varepsilon \beta \rho\right)\phi'' + \left(2\rho + \varepsilon \beta \right) \phi' - U_{*} \phi =0,
	\label{eq: near-horizon master ode}
\end{equation}
where $U_{*} := 2(U/f'')|_{r_+}$.
The solutions are Legendre's functions
\begin{equation}
	\phi = A P_{\alpha}(z) + B Q_{\alpha}(z),
	\label{eq: non-extremal solution}
\end{equation}
with integration constants $A$ and $B$, and
\begin{equation}
	\alpha = \frac{1}{2}\left(- 1 + \sqrt{1 + 4U_{*}}\right),\quad z = 1 + \frac{2\rho}{\varepsilon\beta}.
\end{equation}
The combination $\rho/\varepsilon$ in the expression for $z$ is symptomatic of the tension between the extremal and near-horizon expansions.
In the strictly sub-extremal case, the near-horizon limit corresponds to $z \rightarrow 1$, and the non-extremal solution \eqref{eq: non-extremal solution} takes the form
\begin{equation}
	\phi \sim A \left[1+ \frac{1}{2}\alpha(\alpha-1)(z-1) + \dots \right] + B \left[c-\frac{1}{2}\log(z-1) + \dots \right],
	\label{eq: non-extremal solution nh}
\end{equation}
for some constant $c$ involving digamma functions.
Finite boundary conditions at the horizon require $B=0$, and the remaining solution is regular on the horizon.
We will discard the former branch of solutions even in the extremal case, regardless of the behaviour at the boundaries, as they would not arise from regular non-extremal solutions.

\subsubsection{Extremal Limit}

The horizon ceases to be analytic in the extremal limit as $z$ becomes large.
Setting $B = 0$ from the outset for finite boundary conditions at the horizon in the sub-extremal case, the asymptotic expansion of the solution \eqref{eq: non-extremal solution} in $\varepsilon$ is
\begin{equation}
	\begin{aligned}	
		\phi \sim & A \left[\frac{\Gamma(-1-2\tilde{\alpha})}{\Gamma(-\tilde{\alpha})^{2}}\left(\frac{\beta\varepsilon}{\rho}\right)^{1+\tilde{\alpha}} + \dots + \frac{\Gamma(1+2\tilde{\alpha})}{\Gamma(\tilde{\alpha})^{2}}\left(\frac{\beta\varepsilon}{\rho}\right)^{-\tilde{\alpha}} + \dots \right],
	\end{aligned}
	\label{eq: near-extremal solution}
\end{equation}
where $\tilde{\alpha}$ is the extremal limit of $\alpha$.
The subleading terms come in the form of a power series in $\varepsilon/\rho$, so even for $\tilde{\alpha} \in \mathbb{C}$ we can simply rescale the integration constants by the most negative real power of $\varepsilon$ respectively in order to obtain finite solutions in the extremal limit.
What remains is
\begin{equation}
	\phi = \phi_{0} \rho^{\tilde{\alpha}},
	\label{eq: extremal solution}
\end{equation}
where $\phi_{0}$ is some constant.
Had we not set $B=0$ from the outset, the other solution with exponent $-(1+\tilde{\alpha})$ could have also survived.
Note that these are precisely the deformations from \cite{Horowitz:2022mly, Horowitz:2023xyl} of the form \eqref{eq: critical scaling}, obtained by directly solving the extremal limit of the master equation \eqref{eq: master ode}
\begin{equation}
	\rho^{2}\phi'' + 2\rho \phi' - \tilde{U}_{*} \phi = 0\,,
	\label{eq: hks equation}
\end{equation}
with $\tilde{U}_{*}$ the extremal limit of $U_{*}$.
In terms of their notation, the scaling exponents are
\begin{equation}
	\gamma_{\pm} = -\frac{1}{2} \pm \left(\tilde{\alpha}+\frac{1}{2}\right) = \frac{1}{2}\left(-1 \pm \sqrt{1 + 4\tilde{U}_{*}}\right).
	\label{eq: hks solution}
\end{equation}

We can understand the critical scaling of the near-horizon solutions from a holographic perspective \cite{Gralla:2018xzo}.
The near-horizon limit of the extremal AdS RN metric
\begin{equation}
	\dd s^{2} = \frac{2}{f''(r_{H})}\left[-\rho^{2} \left(\frac{f''(r_{H})}{2}\dd t\right)^{2} + \frac{\dd \rho^{2}}{\rho^{2}}\right] + r_{H}^{2} \dd \Omega_{d}^{2}\,,
\end{equation}
is the Robinson-Bertotti solution, \textit{i.e.} $\mathrm{AdS}_{2} \times S^{d}$ with $\mathrm{AdS}_{2}$ length scale $L_{2}=\sqrt{2/f''(r_{H})}$.
We expect this $\mathrm{AdS}_{2}$-factor to be generic (under suitable assumptions) to the near-horizon geometry of extremal BHs even in theories with other matter fields and higher-derivative corrections \cite{Kunduri:2007vf, Figueras:2008qh} --- see \cite{Kunduri:2013gce} for a review on this.
This means that we expect to find such critical scaling behaviour of stationary perturbations in the near-horizon limit even in such more general theories.

If we are only interested in finding the scaling dimensions themselves (which is the case), it is more straightforward to follow the procedure outlined in \cite{Horowitz:2022mly}.
By using the isometries of the near-horizon geometry, the problem is reduced to finding $\mathrm{AdS}_{2}\times S^{d}$ backgrounds and decomposing the perturbations with respect to the enhanced isometries thereof --- details for this are given in appendix \ref{app: nh decomposition}.
The resulting perturbations equations are easier to decouple and will just be algebraic (quadratic) in $\gamma$.

In section \ref{app: scalar toy model} we discuss a simple toy model of massive scalar field on AdS RN that mimics the near-horizon deformations of extremal BHs.

\subsubsection{Singularities \label{subsubsec: singularities}}

In real terms, this translates to the following scaling for the perturbed metric and Weyl tensors in Schwarzschild-like coordinates:
\begin{equation}
	h_{\cdot \cdot} \sim \rho^{\gamma}, \quad \delta C_{\cdot \cdot \cdot \cdot} \sim \gamma(\gamma-1) \rho^{\gamma-2}.
	\label{eq: scaling}
\end{equation}
Either by changing coordinates $(t,\rho) \mapsto (t,z)$ with $z=1/\rho$ or by going into an orthonormal frame, one sees that all scalar invariants take the form
\begin{equation}
	S = \rho^{n \gamma},\quad n \in \mathbb{N^{+}},
\end{equation}
no matter the number of covariant derivatives acting on tensors, so long as $S$ is a scalar.
Given a causal geodesic with tangent vector $u$, normalised as $u^{2} \in \{-1,0\}$, one can construct a parallel-propagated orthonormal frame by taking $\hat{e}\indices{_{0}^{\mu}} = u^{\mu}$, so that the component of the Riemann tensor entering the geodesic deviation equation takes the form
\begin{equation}
	\hat{R}\indices{^{1}_{010}} = \left(u^{2}\right)^{2} R_{0101} \supset \left(u^{2}\right)^{2} \rho^{\gamma-2}.
\end{equation}
We therefore see that a \textit{timelike} scalar polynomial (\sp) or parallel-propagated (\pp) singularity, in the sense of \cite{Hawking:1973uf}, occurs when $\gamma<0$ or $0<\gamma<2$ respectively.

At first sight, the presence of either type of singularity on the extremal horizon might be concerning, and prompt a dismissal of the solution.
However, \eqref{eq: non-extremal solution nh} shows that all of these singular solutions arise as non-extremal solutions that are regular at the horizon, so they are ``good" and hence admissible singular solutions in the sense of \cite{Gubser:2000nd}.
For AdS RN BHs, \cite{Horowitz:2022leb} takes the perspective that modes with $\gamma<0$ signal an RG instability in the IR of the dual field theory and that the gravitational theory will non-linearly deform to a less symmetric near-horizon geometry.

In physical terms, \pp--singularities cause divergent tidal forces for point-particle observers --- in other words, they lead to breakdown of the {\bf worldline-EFT}, as argued in \cite{Horowitz:2024dch}.
In and of itself this is not necessarily problematic since it may only point to a failure of the geodesic/geometric optics approximation and what is more relevant is whether wavepackets or more precisely quantum field evolution can be made sense of past the singularity.
It is also worth noting that for certain singularities where the worldline-EFT breaks down, the propagation of strings may nevertheless be well defined \cite{deVega:1990ke,Tolley:2005us} for $\gamma>1$.

By contrast, \sp--singularities (if they exist) are clearly dangerous even for quantum fields, as they would seem to undermine the validity of the EFT.
One crucial difference between the two types of singularities is that for \pp--singularities the metric perturbation remains small as we approach the horizon, meaning we expect their properties to be robust predictions valid even in the non-linear theory.
For the would-be \sp--singularities, metric perturbations blow up and it would be necessary to understand the non-linear dynamics to understand the fate of the (potential) singularity.
From the holographic perspective, this corresponds to thinking about how quantities behave as the temperature of the dual theory $T \rightarrow 0$.
Although the tidal forces become large as $T \rightarrow 0$, this behaviour is mild in comparison to a potential \sp--singularity.

In what follows we shall focus principally on $\gcrit = 0$ as the critical value that divides regular or mildly \pp--singular situations, for which metric perturbation theory is under control, from the potentially truly singular behaviour, for which metric perturbation theory breaks down.
We certainly do not preclude the possibility that breakdowns of descriptions also occur in other cases, and in particular for $0<\gamma<2$, viewing for instance $\gcrit=2$ or $\gcrit=1/2$ as the marginal cases.
However the behaviour one can identify for $\gamma<0$ does signal a problematic class of singularities on which we will focus.

\section{Massive Scalar EFT on AdS RN \label{app: scalar toy model}}

The deformations of interest in the following section on UV sensitivity are gravitational \textit{scalar} modes. It is thus natural to expect that we can obtain similar results with a scalar field EFT which is decoupled from the scalar gravitational perturbations\footnote{While we do expect  the behavior of the scalar gravitational mode to be captured by a particular class of scalar field EFTs, not all scalar field EFTs are representative of the gravitational scalar mode. This scalar example only serves as simple illustration on how an EFT can correctly capture the behaviour of the UV even at the onset of an apparent singularity and how perturbations breakdown before the EFT does, indicating a lack of evidence for high UV sensitivity. However we do not expect all parameters of this scalar EFT to be representative of a fully-fledged gravitational model.}. In this section we illustrate this with a concrete EFT example of a scalar EFT endowed with a (partial) UV completion.

\subsection{Spectator Scalar non-minimally coupled to Einstein-Maxwell}
Consider a massive scalar field $\phi$ non-minimally coupled to Einstein-Maxwell theory \eqref{eq: em action}. We shall consider the scalar field as a spectator so that its background solution is $\phi=0$ meaning we can neglect its self-interactions in the analysis of perturbations and that it decouples from the gravitational perturbations.
Nevertheless, the scalar can still interact with the background curvature and Maxwell field with the leading operators being given by
\ba
	\kappa \mathcal{L}_{\phi} &=& - \frac{1}{2}\left(g^{\mu\nu}-\frac{\beta_1}{\Lambda^2} R^{\mu\nu} -  \frac{\beta_2}{\Lambda^2} R g^{\mu\nu}  - \frac{\beta_3}{\Lambda^2} \widehat F^{\mu \alpha}\widehat F^{\mu}{}_{\alpha}   -  \frac{\beta_4}{\Lambda^2} \widehat F^2 g^{\mu\nu}      \right)(\nabla_{\mu}\phi)(\nabla_{\nu}\phi)  \nonumber \\
	&&  - \frac{1}{2} m^{2}\phi^{2}\(1-\frac{\beta_5}{\Lambda^2}   \widehat F^2 \) {+\ldots} \,,
	\label{eq: scalar toy model}
\ea
in a generic EFT. We have not included $R \phi^2$ interactions as including them would mean we are no longer considering the IR theory to be Einstein-Maxwell and a decoupled scalar, and in addition they can easily be removed by a field redefinition. Due to the shift symmetry recovered in the massless limit we regard terms such as $R^2 \phi^2$ as arising at order $m^2/\Lambda^4$ meaning they can be neglected in our analysis. The particular interaction $\beta_1$ arises at loop level \cite{Hollowood:2016ryc} from self-interactions of the scalar, and the $\beta_3$--and $\beta_4$--interactions can arise from loops of charged particles which couple to $\phi$.

When this theory is coupled to dynamical gravity, or more precisely Einstein-Maxwell theory \eqref{eq: em action}, we may remove the $\beta_1$, $\beta_2$ couplings by a field redefinition,  where they generate an additional contribution to the prototypical Goldstone EFT operator $\mathcal{L} \supset (\nabla \phi)^{4}$ as well as additional $F^2 \nabla \phi \nabla \phi $ interactions\footnote{Additionally, we need to perform wavefunction and mass renormalizations due to the cosmological constant.}. Ignoring the self-interactions of the scalar, the relevant terms are therefore
\begin{equation}
	\kappa \mathcal{L}_{\phi} = - \frac{1}{2}\left(g^{\mu\nu}- \frac{\tilde \beta_3}{\Lambda^2} \widehat F^{\mu \alpha}\widehat F^{\nu}{}_{\alpha}   -  \frac{\tilde \beta_4}{\Lambda^2} \widehat F^2 g^{\mu\nu}      \right)(\nabla_{\mu}\phi)(\nabla_{\nu}\phi) - \frac{1}{2} m^{2}\phi^{2}\(1-\frac{\beta_5}{\Lambda^2}   \widehat F^2 \)\phi^{2} \, ,
	\label{eq: scalar toy model_2}
\end{equation}
where the field redefinition invariant coefficients $\tilde{\beta}_{3}$ and $\tilde{\beta}_{4}$ are given in appendix~\ref{app: scalar}.
We also discuss causality/unitarity implications on the respective signs of these coefficients.
In particular, we show how $\tilde{\beta}_4-\beta_5<0$ may be favoured in some marginal cases.

\subsection{Critical Scaling $\gamma_0=0$ \label{subsec: scalar toy model critical scaling}}

The action for the scalar field in the basis \eqref{eq: scalar toy model_2} on the background of the AdS RN solution is
\ba
S &=& \int \d v \int \d^{D-2} \Omega \int \d r r^{D-2} \left\{  \frac{1}{2}\left[- 2 \frac{\partial \phi}{\partial v}\frac{\partial \phi}{\partial r}  - f  \(\frac{\partial \phi}{\partial r}\)^2  \right]\(1+(\tilde \beta_3+2 \tilde \beta_4) \frac{(\psi')^2}{ \Lambda^2}\) \right. \nonumber \\
&& \left. -\frac{1}{2 r^2} (\nabla_{S^d} \phi)^2 \(1+2 \tilde \beta_4\frac{(\psi')^2}{ \Lambda^2}\) -\frac{1}{2}m^2\(1+2 \beta_5\frac{(\psi')^2}{ \Lambda^2}\)  \phi^2 \right\} \, ,
\ea%
with $\psi = \Psi \sqrt{\kappa}$ and $A_{\mu} \d x^{\mu} = - \Psi(r) \d t $. For staticity and spherical symmetry, we decompose the scalar field as
\begin{equation}
 \phi(v,r,x^{i}) = \phi(r) \mathbb{S}(x^{i}).
\end{equation}
For a given multipole $\ell$, in the near-horizon limit $r \rightarrow r_{H}$, the static modes then satisfy
\begin{equation}
	\label{eq: scalar toy model effective equation}
		\frac{(D-3)^2 \sigma}{r_H^2} \(\rho^2 \phi''+2 \rho \phi' \)- \frac{k_S^2}{r_H^2} \frac{ 1+2 \tilde \beta_4 \lambda  }{ 1+(\tilde \beta_3+2 \tilde \beta_4) \lambda }\phi -\frac{m^2(1+2 \beta_5 \lambda)}{ 1+(\tilde \beta_3+2 \tilde \beta_4) \lambda }  \phi =0 \, ,
	\end{equation}
with $k_S^2=\ell(\ell+D-3)$ and
\be \label{lambda}
\lambda := \frac{(\psi')^2}{ \Lambda^2}=\frac{(D-2) }{r_H^2 \Lambda^2} \left[(D-3)+(D-1)\varsigma\right] \, .
\ee
We see that, as expected, the equation for the scalar field \eqref{eq: scalar toy model effective equation} takes the form of \eqref{eq: hks equation} and so the solutions take the form of scaling solutions \eqref{eq: critical scaling} with critical exponent
\begin{equation}
	\gamma = \frac{1}{2} \left(-1 +\sqrt{1+ \frac{4}{(D-3)^{2}\sigma}\frac{m^2(1+2 \beta_5 \lambda)r_{H}^{2}+k_S^2  \(1+2 \tilde \beta_4 \lambda \) }{1+(\tilde \beta_3+2 \tilde \beta_4) \lambda }} \right).
	\label{eq: scalar field critical exponent}
\end{equation}
Since we are working with a truncated EFT action, this formula is only meaningful to first order in $\lambda$, so we write
\begin{equation}
	\gamma = \gamma_{0} + \gamma_{\text{EFT}},\quad \gamma_{0} := \gamma\big|_{\lambda=0} \, ,
\end{equation}
then
\begin{equation}
	\gamma_0 = \frac{1}{2} \left(-1 +\sqrt{1+ \frac{4}{(D-3)^{2} \sigma}(m^{2}r_{H}^{2}+k_S^2) }\right)\,,
	\label{eq: scalar field critical exponent_0}
\end{equation}
is the analogue of what we referred to as $\ggr$ in the introduction. Since we are considering $D$-dimensional geometries which are asymptotically $\mathrm{AdS}_D$, we may consider negative $m^2$ provided we satisfy the $D$-dimensional BF bound $m^2>-\frac{(D-1)^2}{4 L^2}$.
We may therefore make the mode marginal by choosing the BH mass to be
\begin{equation}
	m^2 r_{H}^{2} = -k_{S}^2 \, .
	\end{equation}
	This requires considering BHs of size comparable or larger than the asymptotic AdS$^D$ length scale.
In this case
\begin{equation}
	\gmarg := \gamma\big|_{\gamma_{0}=0} =  2 (\tilde \beta_4 -\beta_5)\frac{1}{\Lambda^2 r_H^2} \frac{(D-2)\left[D-3+(D-1)  \varsigma\right]k_{S}^2}{(D-3)^{2}\sigma}.
\end{equation}
So for this choice we see that the sign of $\gamma$ for a marginal mode is uniquely determined by the combination of Wilson coefficient $ \tilde \beta_4 -\beta_5$.
This simple example illustrates the essential feature noted in \cite{Horowitz:2023xyl,Horowitz:2024dch}, that the nature of the singularity is strongly sensitive to the signs of EFT Wilson coefficients.
For $ \tilde \beta_4 -\beta_5>0$, $\hat{\gamma}$ is positive but generically $\hat{\gamma}<2$ --- this corresponds to the \pp--singularity in the gravitational case.
For $ \tilde \beta_4 -\beta_5<0$, $\gamma$ is negative which corresponds to a potential \sp--singularity.
Indeed, the scalar perturbations diverge at the horizon and generic higher derivative scalar EFT operators also diverge.
At first sight, this feature may seem to indicate a breakdown of the low-energy EFT.
However, as we shall see below, this feature is in fact already present in the (partial) UV completion we shall consider below.
Hence, rather than signalling an EFT breakdown, the EFT is in fact successful at capturing the relevant behaviour of the partial UV precisely.

\subsection{Backreaction and validity of EFT \label{scalarbackreaction}}

Although our scalar example does not couple to gravity at the linear level, it will non-linearly.
Since $\phi \sim \phi_{0} \rho^{\gamma}$, the stress-energy tensor for the scalar scales as
\be
\kappa T \sim r_H^{-2} \phi_{0}^2 \rho^{2 \gamma} \, \quad  \kappa^2 T_{\mu\nu}T^{\mu\nu} \sim r_H^{-4} \phi_{0}^4\rho^{4  \gamma}  \, .
\ee
This will source a gravitational perturbation $h \sim \phi_{0}^2 \rho^{2 \gamma}$.
Thus, if $\gamma<0$ in the scalar sector, this will inevitably lead to metric perturbations going out of control, assuming the EFT remains under control.
Whether or not this is the case depends on the structure of the scalar EFT.
If we assume that i) $\phi$ is a Goldstone-like mode, so that its interactions are dominated by derivative interactions, and ii) the EFT is that of a weakly-coupled UV completion characteristic of string theory as in \eqref{EFTexpansion}, then we expect
\be
\label{EFTexpansion2}
\kappa S_{\rm EFT} = \int \d^D x \sum_{l,m,n,p,q}  c_{l,m,n,p,q} \(\kappa \Lambda^{D-2} \)^l  \,\( \frac{\text{Riemann}}{\Lambda^2}\)^n \( \frac{\nabla}{\Lambda} \)^{2m} \(\frac{\widehat F}{\Lambda}\)^{2 p} \( \frac{\nabla \phi}{\Lambda}\)^{2 q}\, .
\ee
In this case, we expect the scalar EFT to breakdown when $\( \nabla \phi\)^2 \sim \Lambda^2$ which is when $\phi_{0}^2\rho^{2 \gamma} \sim \Lambda^2 r_H^2$. In order to describe the BH within the EFT we require $\Lambda r_H \gg 1$ and so we see that if $\gamma<0$, perturbation theory will necessarily breakdown before the EFT does.

If the scalar EFT is essentially non-gravitational,
the operator expansion can simply be generalized from the flat space $\kappa\to 0$ scalar EFT promoted to a curved a background
\be
\label{EFTexpansion3}
S_{\rm EFT} = \int \d^D x \sum_{m,n,p,q}  c_{m,n,p,q} \,\( \frac{\text{Riemann}}{\Lambda^2}\)^n \( \frac{\nabla}{\Lambda} \)^{2m} \(\frac{ F}{\Lambda}\)^{2 p} \( \frac{\nabla \phi_c}{\Lambda^{D/2}}\)^{2 q}\, ,
\ee
in terms of the canonically normalized field $\phi_c$.
From this expansion, we see that the EFT breaks down when $\( \nabla \phi_c\)^2 \sim \Lambda^D$ or $\( \nabla \phi\)^2 \sim \Lambda^2  \kappa \Lambda^{D-2}$, which occurs when
\be
\phi_{0}^2 \rho^{2 \gamma} \sim \Lambda^2 r_H^2 \( \frac{\Lambda}{\mpl}\)^{D-2} \, .
\ee
If the cutoff $\Lambda$ is well below the Planck scale, then for small BHs the EFT may go out of control before perturbation theory, but for large mass BHs perturbation theory always breakdown before we can reach this conclusion. We cannot regard $\gamma<0$ as indicating UV sensitivity, we can only infer that in this case the fate of the near horizon solution must be dealt with nonlinearly in the full EFT, and only once that nonlinear solution is known can we ascertain if the EFT also breaks down before reaching the horizon.

\subsection{Critical Scaling $\gamma_0=1$ and $\gamma_0=2$ \label{subsec: scalar toy model critical scaling_2}}

In the gravitational case both $\gamma=1$ and $\gamma=2$ are regular since $h$ is analytic in $\rho$.
In our scalar toy model we can achieve this in the absence of EFT corrections by considering a scalar with mass $m$ and tuning the BH mass to be
\be
\sigma =2 f_{\gamma_{\text{crit}}}  \times  (D-3)^2  (m^2 r_H^2+k_S^2)\, ,
\ee
with $f_{1}=1$ or $f_{2}=3$.
When EFT corrections are included, we depart from regularity in general, with marginal scaling exponent given by
\be
\hat{\gamma}\big|_{\gamma_{0}=\gamma_{\text{crit}}}=\gamma_{\text{crit}} -2 f_{\gamma_{\text{crit}}} \lambda (\tilde \beta_3+2 \tilde \beta_4-2 \beta_5) + 2 (\tilde \beta_4-\beta_5) \lambda \frac{k_{S}^2}{(D-3)^{2}\sigma} \, .
\ee

\subsection{Scalar-Maxwell UV completion}
We now explore the extremal BH from a UV perspective. We first provide an explicit example of partial UV completion before pushing the BH to its extreme.

\subsubsection{Example of UV completion}

A partial UV completion of \eqref{eq: scalar toy model} is any theory whose cutoff is parametrically higher than $\Lambda$ which may be used to capture physics at scales where the EFT \eqref{eq: scalar toy model} is certain to break down.
The simplest possibility is to imagine a tree-level UV completion for which new states of mass $\Lambda$ may be integrated in, and the simplest realisation of that is when the new state is a scalar field\footnote{Such a simple example of UV realization is naturally not expected to be representative to the fully fledged gravitational case however our concern here is in the way in which the UV and IR provide an equivalent description of the near horizon physics, and for this purpose this example illustrates well what happens in gravitational situations.}.
With this in mind, consider a heavy state $H$ of mass $\Lambda \gg L^{-1}$ coupled to the photon and the scalar $\phi$, described by the Lagrangian
\be
\label{eq:scalarUV}
 \kappa {\cal L}=\frac{1}{2} R-\frac{1}{2} (\nabla H)^2 - \frac{1}{2} \Lambda^2 H^2 - \frac{1}{4}  e^{\alpha_1 H} \widehat F^2-\frac{1}{2} e^{\alpha_2 H}  (\nabla \phi)^2 -\frac{1}{2}  e^{\alpha_3 H}  m^2 \phi^2  \, .
\ee
Integrating out the heavy field $H$ will generate interactions of the form
\be \label{eq:scalarUVcorrections}
\kappa \Delta {\cal L} = \frac{1}{2 \Lambda^2} \( \frac{1}{4} \alpha_1 \widehat F^2+ \frac{1}{2} \alpha_2 (\nabla \phi)^2 + \frac{1}{2}  \alpha_3 m^2 \phi^2\)^2+\dots\,,
\ee
which corresponds to a contribution
\be
\tilde \beta_4 = \frac{\alpha_1 \alpha_2 }{4} \, , \quad \beta_5=\frac{\alpha_1 \alpha_3}{4} \, .
\ee
The marginal perturbation has $\hat \gamma<0$ when
\be
\alpha_1 (\alpha_2 - \alpha_3)<0 \, .
\ee
At the level of this partial UV completion, we see no obstruction to $\hat \gamma$ being negative or positive assuming the couplings $\alpha_1$ and $\alpha_2-\alpha_3$ can be chosen to have opposite or the same sign.
\subsubsection{Extremality in the UV}

We are now in a position to ask what exactly happens in the (partial) UV theory.
At the level of near-horizon perturbations, the analysis is straightforward.
Since $\widehat F^2$ is constant in the near-horizon limit by symmetry, there continues to be an $\mathrm{AdS}_2\times S^d$ background solution with $\phi=0$ and constant $H=H_0$, as long as
\be \label{H0equation}
H_0 e^{-\alpha_1 H_0} = -\frac{\alpha_1}{4 \Lambda^2} \widehat F^2 \, .
\ee
The decoupled scalar satisfies the equation
\be
\Box \phi - m^2 e^{(\alpha_3-\alpha_2) H_0} \phi=0 \, ,
\ee
which leads to the exact (in perturbation theory) scaling exponent
\begin{equation}
	\gamma_\phi = \frac{1}{2} \left(-1 +\sqrt{1+ \frac{4}{(D-3)^{2} \sigma}\left[e^{(\alpha_3-\alpha_2) H_0}m^{2}r_{H}^{2}+k_S^2 \right] }\right) \,.
\end{equation}
Tuning to the marginal mass $m^2=-k_S^2/r_H^2 $, which to reiterate is allowed by the $D$-dimensional BF bound provided
\be
k_S^2 < \frac{(D-1)^2}{4} \frac{r_H^2}{L^2} \, ,
\ee
 then
\begin{equation}
	\hat{\gamma}_\phi = \frac{1}{2} \left(-1 +\sqrt{1+ \frac{4}{(D-3)^{2} \sigma}k_S^2 \left[1-e^{\frac{(\alpha_3-\alpha_2)}{\alpha_1} \alpha_1 H_0}) \right] }\right) \,.
\end{equation}
Given $\alpha_1 H_0>0$ (as $ \widehat F^2<0$ for electrically charged BHs), $\gamma_\phi $ is manifestly negative (positive) when $\alpha_1 (\alpha_3-\alpha_2) >0$ ($<0$), as expected. We thus see that the sign of the couplings in the UV theory determine the fate of the marginal singular modes, as anticipated by the EFT analysis. Far from breaking down, the EFT is correctly capturing an essential feature of the UV theory.

The new scalar $H$ has perturbations which mix with the scalar photon and graviton modes, but for $(\Lambda r_H)^2 \gg 1$ the mixing is negligible and we have a largely decoupled massive state with
its own scaling exponent for static perturbations
\begin{equation}
	\gamma_H \approx  \frac{1}{2} \left(-1 +\sqrt{1+ \frac{4}{(D-3)^{2} \sigma}(\Lambda^{2}r_{H}^{2}+k_S^2) }\right) \gg 2\,.
\end{equation}
So the perturbation analysis of the UV theory is almost perfectly captured by the EFT, at least for BHs with $\Lambda r_H \gg 1$, as expected.
In the UV theory we can, in principle, consider small BHs with $\Lambda r_H \lesssim 1$.
However, in this particular UV realization, by writing \eqref{H0equation} as
\be
\alpha_1 H_0 e^{-\alpha_1 H_0} = -\frac{\alpha_1^2}{4 \Lambda^2} \widehat F^2 \,,
\ee
we infer
\be
-\widehat F^2 \le \frac{4  e \Lambda^2}{\alpha_1^2}  \, ,
\ee
given $ \widehat F^2<0$ for electrically charged BHs and $x e^{-x} \le e^{-1}$.
This amounts to saying we only have the $\mathrm{AdS}_2\times S^d$ background solution as long as $\Lambda r_H \gtrsim | \alpha_1|$.
For BHs with smaller mass, the heavy field can no longer sit at its minimum and its perturbations will then become significant.

For  $\Lambda r_H \gg 1$, the perturbative analysis remains a good approximation as long as we can neglect the backreaction of the scalar field. Assuming $\alpha_i$ are similar in magnitude, when $(\nabla \phi)^2 $ becomes comparable to $\widehat F^2 \sim r_H^{-2}$ then $H$ starts to receive order unity departures from $H_0$.
Given $\phi = \phi_{0} \rho^{\gamma}$, this occurs precisely when $\phi_{0} \rho^{\gamma} \sim 1$, which is our na\"ive estimate of when perturbation theory breaks down in Sec.~\ref{scalarbackreaction}.
This regime is however still captured by the EFT since it is simply when we need to account for the higher order interactions in \eqref{eq:scalarUVcorrections}. The EFT only begins to go out of control when the perturbative corrections to $H$ are such that the $\alpha_i H$ reach order unity. Assuming all the $\alpha_i$ are comparable in magnitude, this is when
\be
\alpha^2 (\nabla \phi)^2 \sim \Lambda^2 \, ,
\ee
which (apart from the factors of $\alpha$) is consistent with our na\"ive estimate of when the EFT breaks down, namely when
\be
\phi \sim \frac{\Lambda r_H}{|\alpha|} \gg 1 \, .
\ee
Again, this is necessarily after perturbation theory has broken down.

\section{Extremal AdS RN and the UV \label{sec: extremal rn deformations}}

Let us now turn to near-horizon deformations of extremal AdS RN BHs in the EFT of gravity.
In particular, we will closely examine the critical exponents in GR and identify a tower of marginal deformations subject to UV sensitivity, before discussing corrections from extended versions of the examples in the introduction.

\subsection{Critical Exponents \label{subsec: critical exponents}}

The critical exponents for the tensor, vector, and scalar deformations to extremal AdS RN in pure Einstein-Maxwell theory \eqref{eq: em action} are given in \eqref{eq: xrn critical exponents}.
They follow from \eqref{eq: hks solution}, with the effective potentials derived from the spherical decomposition described in appendix \ref{app: sphere decomposition} and explicitly given in \cite{Kodama:2003kk}.
These are functions of the horizon size $r_{H}$ and the multipole moments $\ell \in \{2,3,\dots\}$.

\begin{figure}
	\centering
	\includegraphics[scale=0.75]{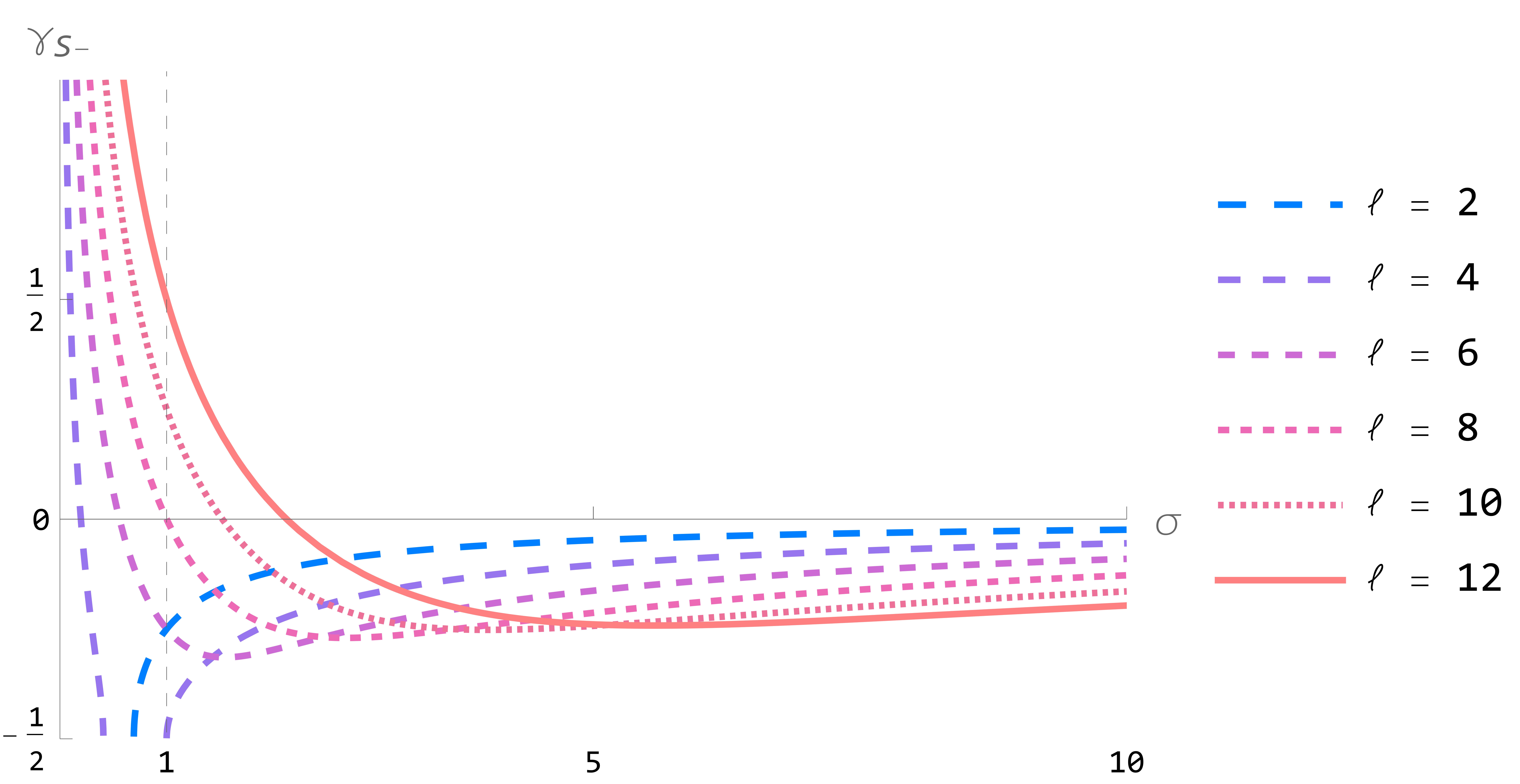}
	\caption{The scaling dimension $\gamma_{S-}$ for gravitational scalar modes in $D=11$ for various $\ell$
 as a function of $\sigma=1+45 r_H^2/32L^2$. Physically, $\sigma>1$ correspond to BHs in AdS, $\sigma<1$ captures BHs in dS, while the line $\sigma=1$ indicates the asymptotically flat limit.
 } \label{fig: scaling fixed dim}
\end{figure}

The tensor, vectors and EM scalar modes all have positive-definite critical exponents, so they are not associated to \sp--singularities on the horizon and do not allow for marginal cases --- we shall therefore not focus on them.
However, the critical exponent for gravitational scalar modes can become arbitrarily small in magnitude and even negative, which is associated to \sp--singularities on the horizon --- this is exacerbated for larger BHs and in higher dimensions. The explicit formula is
\be
\gamma_{S-} = -\frac{1}{2} + \frac{1}{2}\sqrt{5 + \frac{4k_S^2}{(D-3)^{2}\sigma} - 4 \sqrt{1 + \frac{4k_S^2[1+(D-3)\sigma)]}{(D-3)^{2}(D-2)\sigma^{2}}}}.
\ee
In $D=4$, the $\ell=2$ and $\ell=3$ modes are always \pp--singular, and the larger the BH is, the more modes with $\ell > 3$ are also \pp--singular.
Deformations of extremal AdS RN in $D \geq 5$ are always \sp--singular for $2 \leq \ell \leq D-3$ and for sufficiently large BHs with $\ell > D-3$.
In particular, the scaling is marginal with $\gcrit=0$, \textit{i.e.} $\gamma_{S-}=0$ (and hence vulnerable to UV corrections) when
\begin{equation}
	\sigma = -\frac{2}{D-4} + \frac{D-2}{2(D-4)(D-3)^{2}}k_S^2\,.
\end{equation}
This can be achieved for $\ell \geq D-3$ in $D\geq 5$\footnote{Note that $\ell=2$ is only possible when $D=5$.}.
As an example, figure \ref{fig: scaling fixed dim} shows $\gamma_{S-}$ as a function of the BH mass here captured by $\sigma$ given in \eqref{eq:sigma} for several $\ell$-modes in $D=11$.
In the full non-linear theory, generically all modes will be excited, so deformations to extremal AdS RN BHs in $D=4$ and $D>4$ will generically be \pp--and \sp--singular respectively.
This means that deformations to extremal AdS RN BHs are generically singular \cite{Horowitz:2022leb, Horowitz:2022mly}.

Analytic continuation to $\Lambda >0$ is achieved by extending the domain of $\sigma$ from $[1,\infty)$ to $[0,1)$.
Modes with $\ell >D-3$ on backgrounds with sufficiently large BHs are still singular, but for $\ell \leq D-3$ only some BHs are.

The \textbf{asymptotically flat limit} is given by taking $\sigma \rightarrow 1$ and is degenerate with the small BH limit.
As before, the critical exponents are positive-definite for the tensors, vectors, and EM scalars for all physical modes, while for the gravitational scalar modes
\begin{equation}
	\gamma_{S-} = -\frac{1}{2} + \frac{|\ell-\frac{D-3}{2}|}{D-3}.
	\label{eq: asymptotically flat critical exponents}
\end{equation}
Similar to AdS RN, this can be arbitrarily close to zero and even negative.
In $D=4$, we find that $\gamma_{S-} \in \mathbb{N}^{+}$, so the perturbations are smooth on the horizon.
However, in higher dimensions $\gamma_{S-}$ is generally fractional, indicating non-smoothness of the horizon.
For $\ell<3(D-3)$ the modes are \pp--singular, and for $\ell<(D-3)$ they are \sp--singular, with the marginal $\gamma_{S-}=0$ achieved for $\ell=D-3$ which requires $D \ge 5$ to be physical.
Note that for $D > 5$ there is always at least one mode which is \sp--singular at the horizon. For $D=5$ there are no \sp--singular modes, but there is always the marginal mode.

\subsection{Corrections from the UV \label{subsec: uv corrections}}

A notable feature of the gravitational scalar modes is that for every $\ell \geq D-3$ in $D\geq5$, there exists a $r_{H}/L$ (via $\sigma$) for which the near-horizon scaling is marginal, \textit{i.e.} $\gamma_{S-}=\gcrit$.
As previously mentioned, we are particularly interested in the case where $\gcrit=0$, and we see from \eqref{eq: asymptotically flat critical exponents} that the $(\ell,\sigma) = (D-3,1)$-deformation of asymptotically flat RN BHs is always marginal, while for higher $\ell$ it is just a matter of finding a sufficiently large $\sigma$.

This opens up an intriguing possibility for UV sensitivity, generalising the observation first noticed in \cite{Horowitz:2023xyl} for asymptotically flat RN backgrounds --- a similar example was discussed in \cite{Chen:2018jed}.

Of course, generic external sources will not excite modes with specific $\ell$ --- rather they excite a combination of all of them.
Modes with different $\ell$ will also be generated non-linearly, even though these will be suppressed in perturbation theory.
However, for the purpose of finding configurations which are UV sensitive, switching on specific modes is meaningful since it allows us to focus on the special cases where UV sensitivity can be identified.
This is comparable to thinking about scattering in terms of partial wave amplitudes.
Each partial wave amplitude is associated physically with a highly-tuned initial configuration of a spherical partial wave at infinity incoming on the centre.
Although a highly-tuned initial state, it is the correct one to think about the partial wave amplitudes which are the eigenstates of the $S$-matrix in a spherically symmetric scattering problem.
In the same sense we would like to think of the scaling exponent as a field-redefinition invariant observable like the partial wave amplitudes which can be probed by a specific (if highly-tuned) configuration.
Furthermore, in talking about the static deformations, we are already accepting some level of tuning --- truly generic matter perturbations will not lead to static deformations of extremal black holes in the first place, as dynamic modes will dominate the behaviour.

With this in mind, we can now move to the question: What do we actually mean by UV sensitivity?
We generally expect UV completions of GR to correct the action \eqref{eq: em action} by higher-dimension operators including higher-derivative terms as outlined in \eqref{EFTexpansion}.
The effect of these is to change the effective potential (and possibly the effective propagation speed) of decoupled modes in \eqref{eq: master ode}, which show up as a correction to the near-horizon scaling \eqref{eq: critical scaling} that can be \textit{resummed} into the critical scaling exponents into the form of \eqref{eq: critical exponent resummed}.
Schematically, given a collection of higher-derivative operators $\ccO$ with mass dimension $[\ccO]$ supplied with Wilson coefficients $\cO$, we expect
\begin{equation}
	S = S_{\text{EM}} + \int \dd^{D}x \sqrt{-g} \sum_{\ccO} \frac{\cO}{\Lambda^{[\ccO]-D}}\ccO \quad \longrightarrow \quad \geft = \sum_{\ccO}\cO \gamma_{\ccO}.
	\label{eq: critical exponent resummed eft form}
\end{equation}
Let us take $\gmarg$ to denote the scaling exponent in the marginal case, \textit{i.e.} when $\ggr=0$.
The existence of singularities on the horizon of the deformed geometry is very sensitive to the sign of this quantity $\gmarg$ regardless of how small its magnitude is.
This in turn depends on the signs of the coefficients $c_\ccO$, which could in principle depend on the nature or field content of the UV completion --- this is a form of UV sensitivity and is superficially in tension with what we would expect from EFT principles such as decoupling and separation of scales.

\subsubsection{IR Modification: EFT Correction}

As a concrete example of corrections to the critical scaling exponents coming from modifications within the IR and the UV, let us revisit the example from the introduction in more detail.
For these, it is convenient to make full use of the near-horizon geometry symmetries following the procedure in appendix \ref{app: nh decomposition}. We take the action to be
\begin{equation}
	\kappa S =\kappa  S_{\text{EM}} + \frac{ c}{\Lambda^{2}} \int \dd^{5}x \sqrt{-g} (\widehat F_{\mu\nu} \widehat F^{\mu\nu})^{2},
	\label{eq: eft-corrected action}
\end{equation}
where $S_{\text{EM}}$ is given in \eqref{eq: em action}.
It is a specific yet generic correction that one would expect in a low-energy EFT, which is parameterised by the Wilson coefficient $c$.

The theory admits spherically symmetric, static, and electrically charged background solutions of the form \eqref{eq: sss ansatz} with metric functions and electrostatic potential
\begin{equation}
	\label{eq: sss background eft}
	A(r) = B(r) = f(r) - \frac{c}{\Lambda^{2}} \frac{12Q^{4}}{r^{10}},\quad \Psi(r) = \frac{q}{r^{2}} - 16 c \frac{\kappa}{\Lambda^{2}} \frac{q^{3}}{r^{8}},
\end{equation}
and $f(r)$ given in \eqref{eq: ads rn bg}.
The extremal limit is reached when
\begin{equation}
	\label{eq: extremal charge-to-mass eft}
	\frac{Q^{2}}{r_{H}^{4}} = 1+ 2\varsigma + c \frac{48}{\Lambda^{2}r_{H}^{2}}\left(1+2\varsigma\right)^{2},\quad \frac{M}{r_{H}^{2}} = 1 + \frac{3}{2}\varsigma + c \frac{18}{\Lambda^{2}r_{H}^{2}} \left(1+2\varsigma\right)^{2},
\end{equation}
hence
\begin{equation}
	\frac{|Q|}{M} = 1 + \frac{6}{\Lambda^{2}r_{H}^{2}}c
\end{equation}
in the asymptotically flat limit.
This is a special case of the more general result from \cite{Kats:2006xp}, that asymptotically flat RN BHs subject to four-derivative corrections reach extremality when
\begin{equation}
	\frac{|Q|}{M} = 1 + \frac{1}{3 \Lambda^{2}r_{H}^{2}}c_0 \,
		\label{eq: extremal charge-to-mass eft af}
	\end{equation}
in $D=5$, where as we shall see in Section~\ref{sec: uv sensitivity}, $c_0$ is ultimately the field redefinition invariant combination that enters this physical quantity and is given explicitly in terms of the EFT coefficients in \eqref{c0coefficient}.

The fact that the charge-to-mass ratio at which extremality is reached receives corrections has been suggested as providing a resolution or explanation of the WGC \cite{Kats:2006xp,Cheung:2018cwt,Hamada:2018dde}. The idea is that BHs themselves are the objects that allow $Q/M>1$ with violations more pronounced for smaller mass BHs, providing a decay channel by which large charged BHs may radiate charge.
In other words, smaller mass BHs appear as super--extremal {\bf relative} to large mass BHs (even though they are protected by two horizons).
Such an argument would imply  that $c_0>0$ --- in our simple example, this amounts to $c>0$.
More precisely, we may expect this to be true only as long as the BHs are treated purely classically.
Since quantum corrections kick in at $\mathcal{O}(\kappa \Lambda^3)$, we can arguably only require
\be
c>-|{\cal O}(\kappa \Lambda^3)| \, .
\ee
This is precisely what we expect from positivity bounds for gravitationally coupled theories \cite{Alberte:2020jsk,Alberte:2020bdz,Henriksson:2022oeu,deRham:2019ctd,deRham:2020zyh,Alberte:2021dnj,Caron-Huot:2021rmr,Caron-Huot:2022jli,deRham:2022gfe, Caron-Huot:2022ugt,Aoki:2023khq,Hamada:2023cyt,Caron-Huot:2024tsk}.
Separate arguments suggesting (approximate) positivity of corrections to $Q/M$ have been given in \cite{Kats:2006xp,Cheung:2018cwt,Hamada:2018dde}.

As discussed, stationary perturbations of these extremal BHs exhibit critical scaling in the near-horizon region with explicit exponents given in appendix \ref{app: xrn critical exponents}. For the gravitational scalar modes, these take the form of \eqref{eq: critical exponent resummed}, with $\ggr = \gamma_{S-}$  given in  \eqref{eq: xrn critical exponentsSpm}.
In $D=5$ the GR contribution vanishes when $\sigma = -2 + 3k_{S}^{2}/8$, where $k_{S}^{2} = \ell(\ell+2)$, and
hence
\begin{equation}\label{eq: xrn critical exponent marginal_0}
	\gmarg = -\frac{c}{\Lambda^{2} r_{H}^{2}}\frac{72k_{S}^{2}(k_{S}^{2}-4)^{2}}{15k_{S}^{4}-128k_{S}^{2}+256}.
\end{equation}
If $c>0$, then this is negative-definite for all $\ell \geq 2$.
In the asymptotically flat limit, the marginal mode corresponds to $\ell=2$, in which case the expressions above reduce to the ones given in the introduction.
This result can be easily generalised to account for the complete set of four-derivative EFT operators which will be given in Section~\ref{sec: uv sensitivity} as in \eqref{eq: dim-6 corrections action}, leading to
\begin{equation}
	\label{eq: xrn critical exponent marginal}
	\gmarg = -\frac{8 c_0}{3\Lambda^{2} r_{H}^{2}} \, ,
	\end{equation}
where as previously mentioned, $c_0$ is expressed in terms of the other Wilsonian coefficients as in \eqref{c0coefficient}.

\subsubsection{Partial UV Completion: Tower of Massive Spin-0}

Next, consider Einstein-Maxwell-Dilaton theory in $D$ dimensions, with $N$ scalars with masses $m_{i}$ for $i\in\{1,\dots N\}$, described by the following action
\begin{equation}
	\kappa S_{\text{EMD}} = \! \int \dd^{D}x \sqrt{-g} \left[\frac{1}{2}(R-2\Lambda_{\rm c.c.})  -\frac{1}{4}e^{ \sum_{i=1}^{N} \alpha_i \phi_{i}} \widehat F_{\mu\nu} \widehat F^{\mu\nu} + \sum_{i=1}^{N}\( -   \frac{1}{2}\nabla_{\mu}\phi_{i} \nabla^{\mu}\phi_{i} - \frac{1}{2}m_{i}^{2}\phi_{i}^{2} \) \right].
	\label{eq: partial uv completion tower}
\end{equation}
The number of fields can in principle be infinite $N\to \infty$ and the mass spectrum can be taken to be continuous.
We want to treat the $\alpha_i$ perturbatively, and have chosen our normalisation such as to reproduce the Einstein-Maxwell action \eqref{eq: em action} as $m_i \rightarrow \infty$ and the example \eqref{eq: example uv action} from the introduction for $N=1$. This is a partial UV completion to Einstein-Maxwell theory with higher-derivative corrections when the dilaton masses $m_{i}$ are very large relative to the background curvature.
In this case, we can ignore the kinetic term for the dilaton so the individual equations of motion for the dilatons become
\begin{equation}
	m_{i}^{2}\phi_{i} = -\frac{1}{4}\alpha_i \widehat F_{\mu\nu} \widehat F^{\mu\nu} \left[1 + \mathcal{O}\left(\frac{\alpha_i^{2}}{m_{i}^{2}}\right)\right].
\end{equation}
At energies $E \ll m_{i}$, we can find the tree-level effective action by evaluating the full action on the stationary point for the dilaton to give
\begin{equation}
	\kappa S_{\text{EMD}} \supset \int \dd^{D}x \sqrt{-g} \left[ -\frac{1}{4}\widehat F_{\mu\nu} \widehat F^{\mu\nu} + \frac{1}{32\meff^{2}}\left(\widehat F_{\mu\nu}\widehat F^{\mu\nu}\right)^{2} + \dots \right],
	\label{eq: partial uv completion effective action}
\end{equation}
where we have defined the effective mass $\meff$ via
\begin{equation} \label{effectivemass}
	\frac{1}{\meff^{2}} = \sum_{i} \frac{\alpha_i^2}{m_{i}^{2}}.
\end{equation}
We identify this truncated effective theory with the EFT-corrected theory \eqref{eq: eft-corrected action} with Wilson coefficient
\be
c = \frac{ \Lambda^2}{32 \meff^{2}} \, ,
\ee
and the Lorentz-invariant cut-off $\Lambda = {\rm min}\{m_i\} $.
More generally, the low-energy dynamics of the massive dilatons are captured by a $\widehat F^{2(n+1)}$-term at order $\mathcal{O}(\Lambda^{-2n})$ together with higher derivative corrections.

Let us consider $D=5$ again from now on.
The full theory admits $\mathrm{AdS}_{2} \times S^{3}$ background solutions given in \eqref{eq: dilaton background}, and we can use the decomposition \eqref{eq: nh perturbations ansatz} to find the critical scaling exponents.
These take the form of
\begin{equation}
	\gamma = \ggr + \delta \gamma,
	\label{eq: dilaton critical exponent form}
\end{equation}
with $\ggr$ as before and $\delta \gamma$ given in \eqref{eq: xrn critical exponent dilaton} respectively.
The notation $\delta \gamma$ is simply meant to imply that we are working perturbatively in the coupling constants $\alpha_i$, this is however not perturbative in the EFT sense as we are still fully non-perturbatively in $r_H m_i$ and other relevant quantities.
For the marginal case, when the GR contribution vanishes, we now find
\begin{equation}
	\gmarg = -\frac{3k_{S}^{2}(k_{S}^{2}-4)^{2}}{4\left(15k_{S}^{4}-128k_{S}^{2}+256\right)} \left[2\hat{\chi}-\frac{1}{r_{H}^{2}\meff^{2}}\right],\quad \hat{\chi} = \sum_{i=1}^{N} \frac{2 \alpha_i^2}{r_{H}^{2}m_{i}^{2}+k_{S}^{2}}
\end{equation}
which is non-negative when
\begin{equation}
	\meff^{2}r_{H}^{2}\hat{\chi} \leq \frac{1}{2}.
\end{equation}
 From the perspective of the UV theory $\gmarg$ does not have a preferred sign. However if we think of \eqref{eq: partial uv completion effective action} as a partial UV completion for the $F^{4}$-correction, and to be able to trust the BH solutions within the EFT, then we require the dilaton masses to be large relative to the background curvature.
 When we are allowed to integrate out the massive dilaton the leading-order term is
\begin{equation}
	\gmarg = -\frac{1}{r_{H}^{2}\meff^{2}}\frac{9k_{S}^{2}(k_{S}^{2}-4)^{2}}{4(15k_{S}^{4}-128k_{S}^{2}+256)} + \mathcal{O}\left((\meff r_H)^{-4}\right).
	\label{experimental}
\end{equation}
This matches the expectation from the $F^{4}$-contribution in \eqref{eq: xrn critical exponent marginal}, and is manifestly negative.
In this sense, the EFT does have a preferred sign to be embeddable in this class of UV completions.
The marginal case in the asymptotically flat limit corresponds to $\ell=2$, and reproduces the example from the introduction.
What is remarkable in the AdS case is that this is negative for all $\ell$ modes, whenever the BH mass is tuned to make $\ggr$ vanish.
This observation prompts a conjecture that the EFT corrections to marginal perturbations are in fact always negative, or at least bounded from above by a small number.
That this should be the case in gravitational theories is argued for physically in Section~\ref{sec: uv sensitivity} based on the WGC as well as on unitary arguments using holographic considerations and further evidence is given in another situation.

\section{Breakdown of Breakdowns \label{sec: breakdowns}}

In the previous sections, we saw that in the asymptotically AdS case, for every harmonic $\ell$ there exists a BH with tuned size $r_{H}/L$, deformations of the near-horizon geometry which are marginal.
The sign of the marginal scaling exponent $\gmarg$ depends on the signs of the Wilson coefficients, so it seems that the nature of the singularities on the horizon of deformed extremal BHs is strongly sensitive to the details of the UV theory.
This could be seen as suggesting a breakdown of EFT near the horizon, but --- as our previous (partial) UV completion shows --- this is not the case.  This section aims to address this more generally, by determining when metric perturbation theory and EFT break down respectively.
It will indeed turn out that for $\gamma<0$, metric perturbation theory will have gone out of control by the time EFT breaks down.

\subsection{Effective Field Theory \label{subsec: eft breakdown}}

Effective field theory breaks down when curvature invariants of the geometry become large --- this is when the asymptotic expansion in higher-derivative interactions goes out of control.
For this, we have to consider both curvature invariants of the unperturbed and the deformed geometry.

At the level of the unperturbed background geometry, this means that the EFT regime of validity is schematically defined by
\begin{equation}
	\left(\frac{\nabla}{\Lambda}\right)^{a}\left(\frac{\text{Rie}}{\Lambda^{2}}\right)^{b} \ll 1,
	\label{eq: curvature invariants eft regime}
\end{equation}
where again we mean all scalar operators built out of the prescribed number of derivatives and powers of the Riemann tensor.
It can happen that lower order operators are accidentally small, and so to ensure that the asymptotic EFT expansion is meaningful we should impose this criterion in the limit $a,b \rightarrow \infty$. For the sake of concreteness, we can consider a curvature invariant evaluated on the horizon such as
\begin{equation}
	(R^{n}) = R\indices{^{\mu_{1}}_{\mu_{2}}} R\indices{^{\mu_{2}}_{\mu_{3}}} \dots R\indices{^{\mu_{n}}_{\mu_{1}}} \sim \frac{1}{r_{H}^{2n}},
\end{equation}
which leads to the intuitive bound $\Lambda r_{H} \gg 1$ on the background curvature.
There are no other scales in the system, so this is the strongest bound we expect to find on the background.

However, it is also necessary to check that the perturbations to the curvature invariants do not go out of control.
To obtain the strongest bound on $h$, we may consider operators that vanish on the undeformed background, such as
\begin{equation}
	\tilde{R} := R-\bar{R},
\end{equation}
where $\bar{R}$ is the constant background value for the Ricci scalar. At each order in perturbations we may construct the following curvature invariant that is uniquely determined by the linearised perturbations $\tilde{R}^{n} $. Demanding that the perturbations are under control implies
\begin{equation}
	\tilde{R}^{n} \sim \left( \gamma(\gamma+1) \frac{h_{0}\rho^{\gamma}}{L_{2}^2}\right)^{n} \sim  \left( \gamma(\gamma+1) \frac{h_{0}\rho^{\gamma}}{r_H^2}\right)^{n} \ll \Lambda^{2n}\, .
\end{equation}
Treating $\gamma \sim \mathcal{O}(1)$ and denoting $r_{L} = \text{min}\{L_{2},r_{H}\}$, we see that we require from the limit $n \rightarrow \infty$,
\begin{equation}
	h \sim h_{0}\rho^{\gamma} \ll (\Lambda r_{L})^{2} \, ,
\end{equation}
in order for EFT expansion to remain under control.
For the specific marginal mode $\ggr=0$, the regime of validity is enhanced to
\begin{equation}
	h \sim h_{0}\rho^{\hat\gamma} \ll \frac{1}{\hat{\gamma}}(\Lambda r_{L})^{2} \, .
\end{equation}
Assuming, as is expected from positivity bounds \cite{Tolley:2020gtv,Caron-Huot:2020cmc,Caron-Huot:2021rmr,Caron-Huot:2022ugt} that the Wilson coefficients are of order unity, then the EFT corrections to the marginal case ensure that $\hat \gamma \sim (\Lambda r_H)^{-2}$.
Putting this together, the bound on the amplitude of the marginal perturbations to remain within the validity of the EFT is
\begin{equation}
	h \sim h_{0}\rho^{\hat\gamma} \ll (\Lambda r_H)^{4} \, .
\end{equation}
Since the EFT validity of the background requires $\Lambda r_H \gg 1$, then the perturbations of all modes are in the regime of EFT validity whenever $h \lesssim \mathcal{O}(1)$.

On a perturbatively constructed background geometry, the dynamics of fields propagating on the background geometry will also be restricted.
The curvature invariants which reflect this will be built out of covariant derivatives acting on fields in addition to the ones present in \eqref{eq: curvature invariants eft regime} --- the effect of these are captured by the wavevectors $k_{\mu}=(k_{v},k_{\rho},\mathbf{k}^{\flat})$ normalised such that $k^{2}=-m^{2}$.
Therefore, the EFT regime of validity for perturbations propagating on top of the background will be given by
\begin{equation}
	\left(\frac{\nabla}{\Lambda}\right)^{a}\left(\frac{\text{Rie}}{\Lambda^{2}}\right)^{b}\left(\frac{k}{\Lambda}\right)^{c} \ll 1.
\end{equation}
The most stringent bounds on the perturbation $k$ are obtained by contractions for which $(a+2b)/c$ is minimised.
Defining $B_{\mu\nu} = R_{\mu\alpha\nu\beta}k^{\alpha}k^{\beta}$ and splitting $\text{Rie} = \overline{\text{Rie}} + \delta \text{Rie}$, we have to linear order
\begin{equation}
	\begin{aligned}
		\Tr B^{b} &\supset \delta R\indices{^{\mu_{1}}_{\alpha_{1}\mu_{2}\beta_{1}}}k^{\alpha_{1}}k^{\beta_{1}} \bar{R}\indices{^{\mu_{2}}_{\alpha_{2}\mu_{3}\beta_{2}}}k^{\alpha_{2}}k^{\beta_{2}} \dots  \bar{R}\indices{^{\mu_{b}}_{\alpha_{b}\mu_{1}\beta_{b}}}k^{\alpha_{b}}k^{\beta_{b}} \\
		&\sim h_{0} |\mathbf{k}|^{2} k_{v}^{2} \gamma(\gamma-1) \rho^{\gamma-2} r_{H}^{6-4b} \mathbb{S}+ \mathcal{O}\(\rho^{\gamma-1}\),\quad b\geq 2.
	\end{aligned}
\end{equation}
When $b=1$,
\begin{equation}
	\delta R_{\mu\nu} k^{\mu}k^{\nu} \sim h_{0} k_{v}|\mathbf{k}| \gamma r_{H}^{-1} \rho^{\gamma-1} + \dots,
\end{equation}
as the leading order $\mathcal{O}(\rho^{\gamma-2})$-contribution is proportional to $h_{L}$ --- we recognise this as pure-gauge.
However, in the $b \rightarrow \infty$ limit, this recovers the bound we previously found for the background.
One may consider similar contractions with lower $(a+2b)/c$.
For instance, consider $C_{\mu\nu} = (R^{2})_{\alpha\mu\beta\nu}k^{\alpha}k^{\beta}$, with which
\begin{equation}
	\begin{aligned}
		\Tr C B^{n} &\supset \delta (R^{2}) \indices{^{\mu_{1}}_{\alpha_{1}\mu_{2}\beta_{1}}}k^{\alpha_{1}}k^{\beta_{1}} \bar{R}\indices{^{\mu_{2}}_{\alpha_{2}\mu_{3}\beta_{2}}}k^{\alpha_{2}}k^{\beta_{2}} \dots  \bar{R}\indices{^{\mu_{n}}_{\alpha_{n}\mu_{1}\beta_{b-2}}}k^{\alpha_{n}}k^{\beta_{n}} \\
		&\sim \gamma (\gamma-1) h_{0} k_{v}^{2} |\mathbf{k}|^{2n}r_{H}^{-2n-2}\rho^{\gamma-2} \mathbb{S},\quad n = b-2 \geq 1.
	\end{aligned}
\end{equation}
In the limit $b\rightarrow \infty$, this leads to the bound
\begin{equation}
		|\mathbf{k}| \ll \Lambda^{2} r_{H}.
\end{equation}
Next, we may consider contractions of the form
\begin{equation}
	\begin{aligned}
		(k^{\alpha}\nabla_{\alpha})^{a}\delta R &\sim k_{v}^{a} \partial_{\rho} \delta R + \dots \\
		&\sim \frac{\Gamma(\gamma+1)}{\Gamma(\gamma-a+1)} k_{v}^{a}\rho^{\gamma-a}\frac{h_{0}\left[12-3k_{S}^{2}+2\gamma(\gamma+1)(5+\sigma)\right]}{18r_{H}^{2}}\mathbb{S} + \mathcal{O}\left(\rho^{\gamma-a+1}\right),\quad a \geq 2.
	\end{aligned}
\end{equation}
As $a \rightarrow \infty$, this leads to the bound
\begin{equation}
	k_{v} \ll \Lambda^{2} \rho\,,
\end{equation}
for generic $\gamma$.
This is remarkable in and of itself --- it tells us that non-static modes exit the regime of validity of the EFT as we approach the horizon.
At finite temperature, or equivalently for non-extremal black holes, the scaling in $\rho$ is replaced by the temperature, and the upper bound is finite even as we approach the horizon.
However, in the extremal limit, the description breaks down for any perturbation with finite frequency.
Since this occurs for any finite $\Lambda$, this is not a statement about higher-derivative corrections to the Einstein-Hilbert action \textit{per se}.
Take for instance the tidal deformation caused by other extremal black holes in the extremal multi-black hole configurations. 
Even in the absence of intermediate scale $\Lambda$ corrections, loop corrections from gravitons which are always present will induce UV corrections whose counterterms will be of the above form.
In $D >4$, $\gamma$ is generically non-integer, and the above bounds show us that we are unable to describe non-static perturbations at the horizons as soon as we consider quantum corrections.

\subsection{Metric Perturbation Theory \label{subsec: metric pt breakdown}}

Schematically, the expansion of the Einstein-Maxwell action \eqref{eq: em action} in metric perturbations takes the form
\begin{equation}
	\kappa \, S[h] \sim \int \d^{D}x \sqrt{-g} \, h \,\partial^{2} \left[h + h^{2} + \dots \right].
\end{equation}
The quadratic action in perturbations is the part responsible for linearised perturbation theory.
It is clear that perturbation theory breaks down when the action, as an asymptotic expansion in perturbations, is no longer well-defined.
This occurs when higher-order terms in the action dominate, so a na\"ive guess for when control over perturbations is lost is
\begin{equation}
	h \sim h_{0} \rho^{\gamma} \sim \mathcal{O}(1)\,,
	\label{eq: metric perturbation breakdown}
\end{equation}
in an orthonormal frame.

We should be concerned that the actual coefficients of individual terms in the actual action for perturbations \eqref{eq: em action expansion} might vanish --- in the worst case, all higher-order terms vanish and the quadratic term is trivially under control, however we can check that this is trivially not the case, for instance it suffices to explicitly evaluate the Einstein-Maxwell action on-shell up to quartic order.
By treating the spherical harmonics as $\mathcal{O}(1)$ again, the integrands are approximately
\begin{equation}
	\kappa \mathcal{L}_{\text{on-shell}}^{(3)} \sim \frac{h_{0}^{3}\rho^{3\gamma}}{r_{H}^{2}},\quad {\rm and }\quad\kappa  \mathcal{L}_{\text{on-shell}}^{(4)} \sim \frac{h_{0}^{4}\rho^{4\gamma}}{r_{H}^{2}}
\end{equation}
even when $\gamma_{\text{GR}}=0$.
We see that as long as $h_{0}\rho^{\gamma} \ll 1$ we also have $S^{(4)}_{\text{on-shell}} \ll S^{(3)}_{\text{on-shell}}$, confirming our earlier na\"ive estimate.

The threshold for a breakdown of metric perturbation theory will have already been reached when the EFT is out of control.
Finding singular behaviour on the horizon hence only indicates a breakdown of metric perturbation theory.
This suggests that the configuration described above is unstable, so we expect the near-horizon geometry to be replaced by a different solution.
This need not be static or particularly symmetric in any sense, and it is generically not clear that there should even be a static endpoint to this evolution.
This is consistent with the holographic perspective provided in \cite{Horowitz:2022leb}, where new IR geometries stable within a restricted class of perturbations were found, but doubt is cast on more generally stable solutions being found.

This is supported by the examples in section \ref{subsec: uv corrections}, where the deformations appear to suffer from singularities on their horizons in both the IR theory \eqref{eq: eft-corrected action} and its UV completion \eqref{eq: partial uv completion tower}.
The fact that this effect persists even in the UV therefore suggests that the EFT managed to accurately capture the UV physics.
In this sense, the apparent UV sensitivity of $\gmarg$ and the possibility of $\gmarg<0$ are features rather than bugs of the EFT.
From this perspective --- at least based on what occurs around the marginal $\gamma_{\rm GR}=0$ case --- there appears in fact to be no evidence of UV sensitivity that we can see.
In reaching the marginal case, where the contribution from GR vanish and the leading EFT ones would otherwise be left exposed (or amplified), in all the situations we have witnessed, rather than to allow for this possibility, the UV seems to ``protect" itself pushing $\hat \gamma$ to slightly negative values where the symmetric black hole solution can no longer be trusted and is likely replaced non-linearly by a less symmetric one, where the fate of the EFT corrections are yet to be explored.

\section{UV Sensitivity: A conjecture  \label{sec: uv sensitivity}}

We have seen in explicit examples for perturbations of extremal BHs in asymptotically flat or AdS spacetimes, both for gravitational perturbations and pure scalars, it is possible to find BHs for which specific modes are marginal, meaning that their fate is strongly sensitive to UV corrections.
In addition to corrections coming from the $F^{4}$-term, we can also consider corrections to the critical exponents for deformations of AdS RN due to the full set of four-derivative EFT corrections to the Einstein-Maxwell action
\begin{equation}	
	\label{eq: dim-6 corrections action}	
	\begin{aligned}
		\kappa S_{\text{EFT}} &= \int \dd^{5}x \sqrt{-g} \bigg[\frac{c_{1}}{ \Lambda^{2}} R^{2} + \frac{c_{2}}{ \Lambda^{2}}  R_{\mu\nu}R^{\mu\nu} + \frac{c_{3}}{ \Lambda^{2}}  R_{\mu\nu\rho\sigma}R^{\mu\nu\rho\sigma}+\frac{c_{4}}{\Lambda^{2}}  R \widehat F^{2} \\
		&\quad + \frac{c_{5}}{\Lambda^{2}}  R_{\mu\nu}\widehat F\indices{^{\mu}_{\lambda}}\widehat F^{\nu\lambda} + \frac{c_{6}}{\Lambda^{2}}  R_{\mu\nu\rho\sigma}\widehat F^{\mu\nu} \widehat F^{\rho\sigma} + \frac{ c_{7}}{\Lambda^{2}} (\widehat F^{2})^{2} + \frac{ c_{8}}{\Lambda^{2}}  \widehat F\indices{_{\mu}^{\nu}}\widehat F\indices{_{\nu}^{\rho}} \widehat F\indices{_{\rho}^{\sigma}}\widehat F\indices{_{\sigma}^{\mu}}\bigg].
	\end{aligned}
\end{equation}
It was previously noted in \cite{Horowitz:2023xyl} that these lead to deformations of asymptotically flat RN with negative marginal scaling exponent $\gmarg<0$ when the Wilson coefficients respect the signs expected for WGC and \textit{vice versa}.
More specifically, the field redefinition invariant combination of Wilson coefficients
\begin{equation}
\label{c0coefficient}
	c_{0} = \frac{1}{2}c_{1}+\frac{11}{2}c_{2}+\frac{31}{2}c_{3}+3c_{4}+6(c_{5}+c_{6})+9(2c_{7}+c_{8})\,,
\end{equation}
enters both the extremal charge-to-mass ratio \eqref{eq: extremal charge-to-mass eft af} and $\gmarg$ in \eqref{eq: xrn critical exponent marginal} (for $\ell=2$) within the asymptotically flat limit, with the sign satisfying the WGC and related entropy bounds leading to a negative contribution to $\gmarg$ \cite{Kats:2006xp, Cheung:2018cwt}. Our perspective in this rest of this section is that this is not a coincidence.
As discussed at the end of section~\ref{sec: breakdowns}, this could in fact be seen as a way for the UV to ``protect" itself in situations where it could otherwise lead to a well-defined EFT with exposed UV operators (ie where the leading EFT operators would otherwise be amplified).

\subsection{Near-horizon negativity}

There are proposals for the WGC in AdS, but the status of a definite statement is unclear \cite{Harlow:2015lma, Aharony:2021mpc, Aalsma:2021qga}.
It is however notable that in our specific UV completions, there is a link at the level of individual contributions.
UV corrections are seen to enter $\gmarg$ in \eqref{eq: xrn critical exponent marginal_0} of deformed extremal AdS RN with a negative coefficient causing worse behaviour on the horizons of the deformed extremal AdS RN BHs. If we consider the ratio of charge to mass designed to be unity in the absence of EFT corrections
\be
\eta(M)=\frac{Q^2 (1+3 \varsigma/2)^2}{M^2(1+2 \varsigma)}=1+c \frac{12}{\Lambda^2 r_H^2} \frac{ (1+2 \varsigma)}{(1+3 \varsigma/2)} +\dots\, ,
\ee
we see that individual contributions push towards satisfying the WGC in the sense of $\eta(M_1)>\eta(M_2)$ for $M_1<M_2$.

Our perturbation analysis was exclusively focused on the near extremal horizon geometry which is sufficient to determine the scaling exponents $\gamma$. Locality would suggest that the near-horizon physics cannot know if the full, higher-dimensional spacetime is asymptotically AdS or flat.
Let us suppose that the general tenor of the WGC conjecture is true in the above sense, namely that at tree-level $c_0 \gtrsim 0$, and that this similarly implies that $\hat{\gamma}\lesssim0$ for the marginal mode at tree-level.
It is reasonably to expect that there is a purely holographic interpretation of this result in terms of a putative dual theory to the near-horizon ${\rm AdS}_2 \times S^3$ background. Indeed, from the perspective of $\mathrm{AdS}_{2}$ holography, corrections pushing us towards more negative $\gamma$ (such as the ones from known partial UV completions or with Wilson coefficients obeying known bounds), are precisely the ones trying to move further away from the unitarity bound on the $\mathrm{AdS}_{2}$ scaling dimension $\Delta$.\footnote{The fact that the scaling dimensions $\gamma$ contain non-trivial information about physical UV completions should not come as a surprise.
The effective potential for perturbations around backgrounds, in different limits, is also bounded by infrared and quasi-normal mode causality through the scattering time delay and quasi-normal mode frequencies respectively \cite{Chen:2021bvg, Chen:2023rar, Melville:2024zjq}.}
Taken together, these observations suggest the following argument which we dub near-horizon negativity.

\hspace{1cm}\begin{minipage}{0.8\textwidth}
\paragraph{Near-horizon negativity:}
Whenever the bulk gravitational mode is marginal, we expect tree-level corrections to the scaling coefficient to be negative, regardless of the asymptotics of the bulk geometry. If true, then this result will apply to every $\ell$-mode, for which the BH mass is tuned so that the GR mode is marginal. Accounting for the familiar violations of strict positivity due to gravitational interactions \cite{deRham:2019ctd,deRham:2020zyh,Alberte:2020jsk,Caron-Huot:2021rmr,Henriksson:2022oeu}, \textit{i.e.} possible loop corrections to the previous argument, then our conjecture is that for marginal modes in the gravitational sector
\begin{equation}
	\gmarg \leq +  \frac{\mathcal{O}\left(\kappa \Lambda^{D-2}\right)}{\Lambda^2r_H^2}\,.
	\label{eq: nh negativity}
\end{equation}
\end{minipage}\vspace{0.5cm}

We stress that whilst we do not have a rigorous proof of this conjecture, we do have ``experimental evidence", namely the particular partial UV completion \eqref{eq: partial uv completion tower} satisfies this for every $\ell $ mode as seen in \eqref{experimental}. We provide further evidence in the slightly more non-trivial example in Sec.~\ref{furtherevidence} below.

\subsection{Constraints on Four-Derivative Corrections \label{subsec: constraints from nh}}

Let us now study the consequences near-horizon negativity has on the four-derivative corrections to the Einstein-Maxwell action \eqref{eq: dim-6 corrections action}.
For this, we must determine the EFT corrections to the scaling exponent. There is however one subtlety which is not encountered in the asymptotically flat case.
The generic EFT corrections from $\eqref{eq: dim-6 corrections action}$ modify the background solution in such a way that the asymptotic AdS length scale is corrected. It is only meaningful to compare scaling exponents for theories with the same asymptotics. To account for this, we must either account for the redefinition of $L$ when the EFT corrections are included, or work in a form of the EFT action for which there are no corrections to the asymptotic cosmological constant. The latter is most easily done by performing the field redefinitions that put the EFT corrections into the minimal Weyl tensor form from the outset
\begin{equation}	
	\label{eq: dim-6 corrections action2}	
		\kappa S_{\text{EFT}} = \int \dd^{5}x \sqrt{-g} \bigg[ \frac{ c_{3}}{ \Lambda^{2}}  C_{\mu\nu\rho\sigma}C^{\mu\nu\rho\sigma}+\frac{ c_{6}}{\Lambda^{2}}  C_{\mu\nu\rho\sigma}\widehat F^{\mu\nu}\widehat F^{\rho\sigma} + \frac{\tilde c_{7}}{\Lambda^{2}} (\widehat F^{2})^{2} + \frac{\tilde c_{8}}{\Lambda^{2}}  \widehat F\indices{_{\mu}^{\nu}} \widehat F\indices{_{\nu}^{\rho}} \widehat F\indices{_{\rho}^{\sigma}}\widehat F\indices{_{\sigma}^{\mu}}\bigg],
\end{equation}
where we have defined the field redefinition invariant combinations $c_3$, $c_6$, together with
\begin{eqnarray}
&& \tilde c_7 = c_7+\frac{1}{36} c_1-\frac{7}{36} c_2-\frac{19}{72} c_3+\frac{1}{6}c_4 -\frac{1}{6} c_5-\frac{1}{4} c_6 \, ,\\
&& \tilde c_8=c_8+c_2+\frac{4}{3} c_3+c_5+\frac{4}{3} c_6 \, .
\end{eqnarray}
The corrections to the critical scaling exponents in this theory can then be determined from the general expressions given in appendix \ref{subapp: xrn critical exponents eft}.
From around \eqref{eq: xrn critical exponent dim-6} it can be shown that the marginal critical exponent takes the form of
\begin{equation} \label{finalresult}
	\gmarg = -\frac{1}{\Lambda^{2}r_{H}^{2}}\frac{64k_{S}^{2}}{256-128k_{S}^{2}+15k_{S}^{4}}\tilde{c}_{0},
\end{equation}
and
\begin{equation}
	\begin{aligned}
		\tilde{c}_{0} =& \frac{1}{16} \left(k_{S}^{2}-4\right)^{2} c_{0} + \frac{1}{64}\left(k_{S}^{2}-8\right)\left[\left(72-21k_{S}^{2}\right)c_{3}+8\left(k_{S}^{2}-4\)c_{6}\right]\, .
	\end{aligned}
\end{equation}
Since $\tilde{c}_{0}$ enters $\gmarg$ with a negative coefficient, near-horizon negativity would require $\tilde{c}_{0} \gtrsim 0$.
The asymptotically flat limit corresponds to $k_{S}^{2}=8$ and reproduces the original WGC bound
\begin{equation}
	\tilde{c}_{0}(\ell=2)=  c_0 \gtrsim 0\,.
\end{equation}
The general bound $\tilde{c}_{0} \gtrsim 0$ on AdS RN will give an infinite tower of additional bounds.
Concretely, the new bounds beyond the asymptotically flat WGC come from $\ell > 2$.

In particular we note that if we assume near-horizon negativity is valid in the limit $k_S \rightarrow \infty$, meaning for large BHs, then we infer
\begin{equation}
\hat c_0 = \lim_{\ell \rightarrow \infty} \frac{\tilde{c}_{0}(\ell)}{k_S^4} =  \frac{1}{64}\left(4c_{0}-21c_{3} + 8 c_{6}\right) \gtrsim
-|\mathcal{O}(\kappa \Lambda^{D-2})| \, .
\end{equation}
This conjectured positivity bound is quite independent of the asymptotically flat case on $c_0$.

\subsection{Further Evidence: Einstein-Maxwell-Dilaton} \label{furtherevidence}

We have so far considered the IR gravitational theory to be Einstein-Maxwell plus EFT corrections with \eqref{eq: partial uv completion tower} as an explicit example UV completion. However an obvious extension is to consider the theory \eqref{eq: partial uv completion tower} and split the $N=N_L+N_H$ scalar fields into $N_L$ light dilatons and $N_H$ heavy ones which may be integrated out. The extended IR theory is now a particular Einstein-Maxwell-(Multi-)Dilaton theory
\begin{equation}
	\kappa S_{\text{IR}} = \int \dd^{D}x \sqrt{-g} \left[\frac{1}{2}(R-2\Lambda_{\rm c.c.})  -\frac{1}{4}e^{ \sum_{i=1}^{N_L} \alpha_i \phi_{i}} \widehat F_{\mu\nu} \widehat F^{\mu\nu} + \sum_{i=1}^{N_L}\( -   \frac{1}{2}(\partial \phi_i)^2 - \frac{1}{2}m_{i}^{2}\phi_{i}^{2} \) \right]\,,
	\label{eq: extended IR}
\end{equation}
which itself receives EFT corrections from integrating out the heavy states. The scaling exponents can again be split into
\be
\gamma = \gamma_{\rm IR}+\gamma_{\rm EFT} \, .
\ee
Since the dilaton affects the scaling exponents $ \gamma_{\rm IR}$ at order $\alpha_i^2$, it is clear that the condition for marginality is modified in a manner that depends on $\alpha_i$ and $m_i$ for $i=1 \dots N_L$. Nevertheless it remains the case that we can tune the BH mass such that $ \gamma_{\rm IR}=0$ for a given $\ell$. Once this choice is made, the remaining correction $\gamma_{\rm EFT} $ comes entirely from the heavy states $i=(N_L+1) \dots N$. To order $\alpha_i^2$ this correction is determined from \eqref{eq: xrn critical exponent dilaton}, which keeping only the leading terms in an expansion in $1/m_i^2$ for the heavy states gives
\ba
	\label{eq: xrn critical exponent dilaton_2}
	\gamma_{\rm EFT} =\mathcal{A} \Delta \sum_{i=N_L+1}^{N}  \frac{\alpha_i^2}{m_i^2 r_H^2}
\ea
with $z=1 + \frac{k_{S}^{2}}{4\sigma} - \sqrt{1 + \frac{1+2\sigma}{3\sigma^{2}}k_{S}^{2}}$ and with the overall coefficient $\mathcal{A}$ defined as
\ba
\mathcal{A} &=& -18 k_{S}^{2}(1+2\sigma)(-4z\sigma-4+k_{S}^{2})
+6(1+2\sigma)^{2}\left(-12(z-2)z\sigma+k_{S}^{2}(1+3z)\right) \\
&=& - 6 k_S^2 \frac{(1+2 \sigma)}{\sigma} \( 4(1-\sigma^2)+3 \sigma(2 \sigma-1) \sqrt{1 + \frac{1+2\sigma}{3\sigma^{2}}k_{S}^{2}} \)\,.
\ea
Given $\sigma > 1$ and $ \sqrt{1 + \frac{1+2\sigma}{3\sigma^{2}}k_{S}^{2}}>1$ we infer that $\mathcal{A}$ is always negative,
\ba
\mathcal{A} &<&-6 k_S^2 \frac{(1+2 \sigma)}{\sigma} \( 4(1-\sigma^2)+3 \sigma(2 \sigma-1)\)\\
&<&-6 k_S^2 \frac{(1+2 \sigma)}{\sigma} \( 3+2( \sigma-1)+2 (\sigma-1)^2\)<0 \, .
\ea
A slightly more involved analysis shows that for $\ell \ge 2$, the prefactor $\Delta$ is positive for all $\sigma > 1$. Putting this together we infer that
\be
\hat\gamma <0  \, , \quad  \forall \ \ell \ \ge 2 \, , \  r_H/L \ge 0 \, .
\ee
Thus even though we have modified the IR theory to include a set of light dilatons of arbitrary {\it light} masses, and the condition for marginality is modified, the EFT corrections coming from the heavy states which are sufficiently massive to integrate out are universally negative, consistent with the near-horizon negativity conjecture. Of course we have only demonstrated that this holds in this particular UV completion (at order $\alpha^2)$, and it would be interesting to explore other examples which probe different EFT operators.

\section{Concluding Remarks \label{sec: conclusion}}

In this work, we considered stationary deformations to the near-horizon geometry of extremal BHs in AdS.
We think of these as background tidal deformations induced by matter sources external to the BH. These tidal deformations exhibit critical scaling near the extremal horizon, with critical exponents $\gamma$ given by the scaling dimensions of dual operators in the putative boundary of the near-horizon $\mathrm{AdS}_{2}$ factor.
Due to the rigidity of this near-horizon geometry, this applies universally to generic EFTs in the presence of generic higher-derivative corrections. These higher dimension operators show up as corrections to the near-horizon perturbations that can be resummed into the critical exponents.
Metric perturbations are finite or non-analytic, or appear to be divergent at the horizon, depending on the sign and magnitude of $\gamma$.
We have argued that these singular extremal deformations arise as extremal limits of regular non-extremal solutions, so they constitute physical singular extremal geometries that are admissible in the sense of \cite{Gubser:2000nd}.

These essential features are exhibited by a rather simple scalar field effective theory. We show that the scaling exponents of the scalar field may be tuned to be arbitrary close to zero leading to an apparent UV sensitivity in that the sign of the scaling exponents is uniquely determined by higher dimension EFT operators. To clarify this we explore one particular (partial) UV completion of this theory. An analysis of both the UV and IR theory show that UV theory exhibits all of the same features as the EFT and that the EFT far from going out of control does an excellent job of capturing the essential physics regardless of the scaling exponents.

In the case of Einstein-Maxwell theory with a negative cosmological constant, we identified a discrete tower of marginal modes in the gravitational scalar sector, for which the leading-order contribution to the critical exponents vanishes and EFT corrections dominate, as in the scalar example.
In examples of EFT corrections with Wilson coefficients derived from explicit UV completions or satisfying known bounds, we saw that corrections lead to worse behaviour at the horizon.
We showed that when $\gamma<0$ these singularities are associated to metric perturbation theory breaking down rather than the EFT breaking down, and demonstrated that as in the scalar field example, explicit UV completions have the same property confirming that this is not a failure of the EFT.

Motivated by these explicit examples, we conjecture a set of speculative bounds motivated by the WGC, that UV corrections to marginal perturbations lead to universality more singular behaviour on the horizon. These near-horizon negativity bounds are strictly stronger than the WGC. We demonstrated that this conjecture holds for a more general class of Einstein-Maxwell-Dilaton effective field theories, and argue that it would allow for the UV to ``protect" itself against UV sensitivity, or in other words against situations in which EFT operators become ``exposed" or physically ``amplified".

Within the context of this current work, near-horizon negativity remains a speculative conjecture. To explore it further, it would be instructive to identify other situations, for example by considering loop level UV completions or other more general UV completions and ascertain if more concrete physical arguments can be made more robust. Alternatively, it would be enlightening to explore the existence of any gravitational counter-example (where the scalar mode genuinely starts its life as a gravitational degree of freedom in a UV embeddable theory) and better establish what central physical properties are responsible for the change in behavior.

\acknowledgments

It is our pleasure to thank Mariana Carrillo Gonz\'alez and Sumer Jaitly for initial collaboration, and we would like to express gratitude to Cliff Burgess, Scott Melville, and Maria J. Rodriguez, and especially Jerome Gauntlett, Harvey S. Reall, Akihiro Ishibashi and Toby Wiseman for stimulating and useful discussions. The work of CdR and AJT is supported by STFC Consolidated Grant ST/X000575/1. CdR is also supported by a Simons Investigator award 690508.  CYRC is supported by the Imperial College London President's PhD Scholarship.

\appendix

\section{Spherical Symmetry Decomposition \label{app: sphere decomposition}}

In this appendix, we briefly summarise the spherical decomposition of the EM and gravitational perturbations in Einstein-Maxwell theory \eqref{eq: em action}, most closely following \cite{Kodama:2003kk} and building on \cite{Kodama:2000fa, Kodama:2003jz}.
We focus on scalar perturbations as only they can lead to marginal deformations.
Explicit expressions for the effective potentials are complicated and of secondary relevance, hence omitted.

\subsection{Scalar Perturbations }

The scalar harmonic $\mathbb{S}$ satisfies
\begin{equation}
	(\hat{\Delta}+k_{S}^{2})\mathbb{S} = 0,
	\label{eq: spherical harmonic} 	
\end{equation}
and we define from it scalar-derived vector and tensor harmonics:
\begin{equation}
	\mathbb{S}_{i}=-\frac{1}{k_{S}}\hat{D}_{i}\mathbb{S},\quad	\mathbb{S}_{ij} = \frac{1}{k_{S}^{2}}\hat{D}_{i}\hat{D}_{j}\mathbb{S}+\frac{1}{D-2}\gamma_{ij}\mathbb{S},
\end{equation}
where
\begin{equation}
	k_{S}^{2} = \ell(\ell+D-3),\quad \ell=0,1,\dots\quad .
\end{equation}
However, $\ell=0$ has constant eigenmodes and therefore corresponds to just a shift in the background parameters.
The $\ell=1$ mode corresponds to $\mathbb{S}_{ij}$ vanishing, and therefore needs to be treated carefully.

We parameterise our scalar metric and field strength perturbations as
\begin{subequations}
	\begin{align}
		&\delta g_{ab} = f_{ab}\mathbb{S},\quad \delta g_{ai} = rf_{a}\mathbb{S}_{i},\quad \delta g_{ij} = 2r^{2}\left(H_{L}\gamma_{ij}\mathbb{S}+H_{T} \mathbb{S}_{ij}\right), \\
		&\delta F_{ab} = -\frac{1}{k_{S}}\epsilon_{ab}\mathbb{S}D_{c}B^{c},\quad \delta F_{ai} = \epsilon_{ab}\mathbb{S}_{i}B^{b},\quad \delta F_{ij} = 0,
	\end{align}
\end{subequations}
and define the gauge-invariant variables by
\begin{subequations}
	\begin{align}
		&F = H_{L} + \frac{1}{D-2}H_{T} + \frac{1}{r}D_{a}r X^{a} \, ,\\
		&F_{ab} = f_{ab} + 2 D_{(a}X_{b)} \, ,
	\end{align}
\end{subequations}
with
\begin{subequations}
	\begin{align}
		&X^{a} = g^{ab}\frac{r}{k_{S}}\left(f_{b} + \frac{r}{k_{S}}D_{b}H_{T}\right) \, ,\\
		&B^{b} = \left(r\mathcal{E}^{b}+k_{S}E_{0}X^{b}\right).
	\end{align}
\end{subequations}
There are two master variables satisfying the master equation \eqref{eq: master eq}, the EM and gravitational scalar modes $\Phi^{+}_{S}$ and $\Phi^{-}_{S}$ respectively, that are related to the gauge-invariant variables above via
\begin{subequations}
	\begin{align}
		&F = a_{-}\Phi^{-}_{S} + b_{-}\partial_{r}\Phi^{-}_{S} + a_{+}\Phi^{+}_{S} + b_{+}\partial_{r}\Phi^{+}_{S}, \\
		&\mathcal{E}_{0} = c_{-}\Phi^{-}_{S} + c_{+}\Phi^{+}_{S}, \\
		&\mathcal{E}_{1} = \tilde{a}_{-}\Phi^{-}_{S} + \tilde{b}_{-}\partial_{r}\Phi^{-}_{S} + \tilde{a}_{+}\Phi^{+}_{S} + \tilde{b}_{+}\partial_{r}\Phi^{+}_{S},
	\end{align}
\end{subequations}
for some constants $a_{\pm}$, $b_{\pm}$, $\tilde{a}_{\pm}$, $\tilde{b}_{\pm}$, and $c_{\pm}$.
For $\ell=1$, only the EM scalar mode is gauge-invariant and dynamical.

\section{Near-Horizon Isometries Decomposition \label{app: nh decomposition}}

In this appendix, we will briefly outline how to compute the critical exponents by starting within the near-horizon limit, following \cite{Horowitz:2022mly}.

\subsection{Background \label{subapp: nh background}}

First consider the background. In ingoing coordinates $(v,r,x^{i})$ with $\dd v = \sqrt{A/B}\, \dd t+ \dd r/B$, the most general static and spherically symmetric Ansatz \eqref{eq: sss ansatz} becomes
\begin{subequations}
	\begin{align}
		\dd s^{2} &= -B \dd v^{2} + 2 \dd v\, \dd r + r^{2} \dd \Omega_{d}^{2}, \\
		F &= \sqrt{\frac{B}{A}} \Psi'(r)\, \dd v \wedge \dd r,
	\end{align}
\end{subequations}
so in the near-horizon limit
\begin{subequations}
	\label{eq: freund-rubin ansatz}
	\begin{align}
		\dd s^{2} &= -\BNH \rho^{2} \dd v^{2} + 2 \dd v\, \dd \rho + r_{H}^{2} \dd \Omega_{d}^{2}, \\
		F &= \sqrt{\frac{\BNH}{\ANH}} \FNH \dd v \wedge \dd \rho,
	\end{align}
\end{subequations}
where $\ANH = \frac{1}{2}A''(r_{H})$, $\BNH = \frac{1}{2}B''(r_{H})$, and $\FNH = \Psi'(r_{H})$ are constants.
This is the form of the $\mathrm{AdS}_{2} \times S^{d}$ Ansatz we will use.
It is given in Gaussian null coordinates, which are a natural and well-defined set of coordinates near null surfaces.

\subsection{Perturbations \label{subapp: nh perturbations}}

Using the isometries of the $\mathrm{AdS}_{2}$ factor in the near-horizon geometry, we can decompose the metric and field strength perturbations in the scalar sector as
\begin{subequations}
	\label{eq: nh perturbations ansatz}
	\begin{align}
		\delta g &= \rho^{\gamma} \bigg[\sqrt{\frac{\ANH}{\BNH^{3}}} f_{00}\, \mathbb{S} \left(\sqrt{\ANH\BNH}\rho^{2} \,\dd v^{2} + 2 \dd v\, \dd \rho + \frac{1}{\sqrt{\ANH\BNH}}\rho^{-2} \dd \rho^{2} \right)\\
		&+ 2 f_{0}\, \mathbb{S}_{i}\, \dd x^{i} \left(\rho \, \dd v\,  + \frac{1}{\sqrt{\ANH\BNH}}\rho^{-1}\dd \rho \right) + (h_{L}\gamma_{ij}\mathbb{S} + h_{T}\mathbb{S}_{ij}) \dd x^{i} \dd x^{j}\bigg] \notag \\
		\delta F &= \rho^{\gamma} \left[a\, \mathbb{S}\, \dd v \wedge \dd \rho + \rho\, e_{0}\, \mathbb{S}_{i}\, \dd v \wedge \dd x^{i} + \rho^{-1} e_{1}\, \mathbb{S}_{i}\, \dd \rho \wedge \dd x^{i}\right],
	\end{align}
\end{subequations}
where $f_{00}$, $f_{0}$, $h_{L}$, $h_{T}$, $a$, $e_{0}$, and $e_{1}$ are constants.
In order for $\delta F$ to be closed, we also need
\begin{equation}
	a = -\frac{\gamma + 1}{k_{S}} e_{0}.
\end{equation}

\subsection{Explicit Solution in $D=5$ \label{subapp: explicit nh solution}}

For concreteness, let us give explicit expressions for the perturbations in Einstein-Maxwell theory \eqref{eq: em action} for $D=5$.
The near-horizon Ansatz solves the background equations of motion for
\begin{equation}
	\ANH= 2\FNH^{2} \kappa \frac{\sigma}{1+2\sigma},\quad \BNH = \frac{4\sigma}{r_{H}^{2}}
\end{equation}
and the perturbation equations are solved by
\begin{subequations}
	\label{eq: explicit perturbations}
	\begin{align}
		h_{L}&=0 \\
		f_{00} &= \frac{h_{T}}{6r_{H}^{4}(\gamma+1)(\gamma+2)} \left(16-3k_{S}^{2}+28\gamma(\gamma+1)\sigma\right) \\
		f_{0} &= \frac{h_{T}}{6k_{S}r_{H}^{2}}\left(-\frac{k_{S}^{2}}{\gamma+1}-12\gamma \sigma\right) \\
		a &= - \frac{\gamma+1}{k_{S}} e_{0} = \frac{k_{S}}{r_{H}^{2}\gamma} e_{1} = \frac{h_{T}}{4\sqrt{2}} \frac{4-k_{S}^{2}+4\gamma(\gamma+1)\sigma}{\kappa r_{H}^{3} \sqrt{1+2\sigma}}\,,
	\end{align}
\end{subequations}
up to gauge transformations.

\section{Massive Scalar Field coupled to a Gauge Field \label{app: scalar}}

In this section we shall provide some of the details related to the massive scalar field $\phi$ which is non-minimally coupled to a Maxwell field, with the leading operators being given by
\ba
	\kappa \mathcal{L}_{\phi} &=& - \frac{1}{2}\left(g^{\mu\nu}-\frac{\beta_1}{\Lambda^2} R^{\mu\nu} -  \frac{\beta_2}{\Lambda^2} R g^{\mu\nu}  - \frac{\beta_3}{\Lambda^2} \widehat F^{\mu \alpha}\widehat F^{\mu}{}_{\alpha}   -  \frac{\beta_4}{\Lambda^2} \widehat F^2 g^{\mu\nu}      \right)(\nabla_{\mu}\phi)(\nabla_{\nu}\phi)  \label{eq: scalar toy model_app} \\
	&&- \frac{1}{2} m^{2}\phi^{2}\( 1- \frac{\beta_5}{\Lambda^2}  \widehat F^2 \) \,,\notag
\ea
as considered in Section~\ref{app: scalar toy model}.
Following field redefinitions as well as wavefunction and mass renormalisations to account for the presence of the cosmological constant, we are led to the following action
\begin{equation}
	\kappa \mathcal{L}_{\phi} = - \frac{1}{2}\left(g^{\mu\nu}- \frac{\tilde \beta_3}{\Lambda^2} \widehat F^{\mu \alpha}\widehat F^{\nu}{}_{\alpha}   -  \frac{\tilde \beta_4}{\Lambda^2} \widehat F^2 g^{\mu\nu}      \right)(\nabla_{\mu}\phi)(\nabla_{\nu}\phi) - \frac{1}{2} m^2\(1-\frac{\beta_5}{\Lambda^2}  \widehat F^2 \)\phi^{2} \, ,
	\label{eq: scalar toy model_2app}
\end{equation}
where scalar quartic self-interactions of the form $(\nabla\phi)^{4}$ are ignored as they are irrelevant on a background where $\langle \phi\rangle=0$.
The two field redefinition invariant quantities $\tilde\beta_{3}$ and $\tilde\beta_{4}$ are given by
\ba
 \tilde \beta_3= \beta_3+\beta_1 \, , \quad {\rm and}\quad
 \tilde \beta_4= \beta_4-\frac{1}{2(D-2)} \beta_1+\frac{(D-4)}{2(D-2)} \beta_2 \, .
\ea
%

If gravity is non-dynamical, there is a preferred sign for $\tilde \beta_3$ based on positivity/causality arguments in Minkowski spacetime. For instance, in the field basis \eqref{eq: scalar toy model_2} we may consider positivity bounds applied to photon-scalar scattering in Minkowski. These imply $\tilde \beta_3>0$, a result which may easily be inferred by considering the propagation of the scalar in the background field of a point charge and demanding that the speed of propagation is subluminal.
When extended to the gravitational case, the positivity bound is expected to be weakened to something of the form
\be
\tilde \beta_3 > - \mathcal{O}\left(\kappa \Lambda^{D-2}\right) \, ,
\label{eq:beta3}
\ee
as discussed in \cite{deRham:2019ctd,deRham:2020zyh,Alberte:2020jsk,Chen:2021bvg,Caron-Huot:2021rmr,Caron-Huot:2022ugt,Chen:2023rar}.
On the other hand, causality does not immediately constrain $\tilde \beta_4$ because it does not change the lightcone structure of the effective metric for the scalar. It does however change the effective mass of fluctuations.
Consider the above theory on the $\mathrm{AdS}_{2} \times S^d$ that describes the near-horizon geometry of an extremal BH.
The fluctuations may be viewed from the perspective of the two-dimensional  $\mathrm{AdS}_{2}$ geometry.
Each $\ell$-multipole fluctuation can be viewed as a two-dimensional scalar of mass $m_{\rm eff}$ which should respect the $D=2$ Breitenlohner-Freedman (BF) bound
\be
m_{\rm eff}^2> -\frac{1}{4 L_{2}^{2}} \, .
\ee
As can be seen from the master equation \eqref{eq: scalar toy model effective equation}, the effective mass is given by
\be
m_{\rm eff}^2=\frac{m^2(1+2 \beta_5 \lambda)}{1+(\tilde \beta_3+2 \tilde \beta_4) \lambda }+\frac{k_S^2}{r_H^2}  \frac{1+2 \tilde \beta_4 \lambda }{1+(\tilde \beta_3+2 \tilde \beta_4) \lambda} \, ,
\ee
with $\lambda$ given in \eqref{lambda}. Demanding that the monopole $\ell=0$ satisfies the 2D BF bound implies
\be
\frac{m^2(1+2 \beta_5 \lambda)}{ (1+(\tilde \beta_3+2 \tilde \beta_4) \lambda )} \ge -\frac{1}{4 L_{2}^2} \, .
\ee
If we consider a scalar of mass $m$ which saturates the BF bound for $\Lambda=\infty$ (\textit{i.e.} $\lambda=0$), then for that same state to respect the bound at finite $\lambda$ we require the denominator of the l.h.s. to be positive.
This suggests the unitarity/positivity bound
\be \label{bound101}
\tilde \beta_3+2 (\tilde \beta_4- \beta_5)\ge 0 \, .
\ee
More generally, the range of masses for which there are two choice of boundary conditions in ${\rm AdS}_2$ is
\be
-\frac{1}{4 L_{2}^{2}} \le m_{\rm eff}^2 \le \frac{3}{4 L_{2}^{2}} \, .
\ee
Making a similar argument for a monopole state which is chosen to saturate the upper bound for $\Lambda=\infty$, \textit{i.e.} $m^2 = 3/(4 L_{2}^{2})$, then to continue to have an interpretation as a state with alternative boundary conditions at finite $\Lambda$, we would require
\be \label{bound102}
\tilde \beta_3+2( \tilde \beta_4 - \beta_5)\le 0 \, .
\ee
As can be seen from these arguments, \eqref{bound101} and \eqref{bound102} are in conflict with each other and so in practise at the level of this analysis, there are no direct known positivity bound on $\tilde \beta_4- \beta_5$. In particular there seems to be no obstruction from having the coefficient $\tilde \beta_4- \beta_5$ negative. In fact
the only case where both \eqref{bound101} and \eqref{bound102} are simultaneously satisfied is when $\tilde \beta_3+2 \tilde \beta_4-2 \beta_5=0$. In this case,
 since  $\tilde{\beta}_3$ is mainly positive (or only at most weakly negative as suggested by \eqref{eq:beta3}), one would expect $\tilde{\beta}_4- \beta_5$ to be mainly negative.

\section{Boundary Conditions \& Multi-Black Hole Solutions \label{app: bcs}}

A stationary perturbation of a RN BH with appropriate boundary conditions at the horizon and asymptotic infinity constitutes a linearisation of a BH solution that is close to the original RN geometry, and should be subject to uniqueness theorems \cite{Emparan:2008eg, Hollands:2012xy}.
More explicitly, the effective potentials and their so-called $S$-deformations given in \cite{Kodama:2003kk} can be used to show that extremal modes that are finite at the horizon are inconsistent with the right fall-off conditions at asymptotic infinity, establishing perturbative uniqueness.

This means that additional sources able to change the asymptotic boundary conditions are required to support near-horizon modes with the desired near-horizon scaling and fall-off conditions at infinity.
We can understand this from the extremal limit of the sub-extremal solution \eqref{eq: near-extremal solution} --- scaling in $\varepsilon$ will generically be different at asymptotic infinity, so the extremal limit will produce non-normalisable or trivial modes in the absence of external sources.
Physically, these deformations should therefore be thought of as sourced by tidal deformations.

An explicit realisation of this are multi-BH solutions, which are known to be exact solutions in Einstein-Maxwell theory.
In practice a convenient way of realising this is to simply impose Dirichlet boundary conditions at finite distance, which is particularly natural in asymptotically AdS spaces.

\subsection*{Multi-Black Hole Solutions}
\label{subapp: majumdar-papapetrou}

The Majumdar-Papapetrou solution is an illustrative example of an explicit non-linear realisation of the deformations above \cite{Myers:1986rx}.
This is an exact solution to the Einstein-Maxwell equations \eqref{eq: einstein eq} and \eqref{eq: maxwell eq} given by the metric
\begin{equation}
	\dd s^{2} = -H^{-2}\dd t^{2} + H^{\frac{2}{D-3}}\eta_{ij}\dd x^{i}\dd x^{j},\quad H(\mathbf{x}) = 1 + \sum_{i=1}^{N} \frac{M_{i}}{|\mathbf{x}-\mathbf{x}_{i}|^{D-3}},
	\label{eq: majumdar-papapetrou}
\end{equation}
and describes a static spacetime with $N$ extremal BHs with masses $M_{i}$ and horizons at $\mathbf{x}_{i}$.
By linearising this exact solution around the horizon of any individual BH, we can see how it is tidally deformed by the other ones.

For convenience, let us pick $\mathbf{x}_{1}=\mathbf{0}$ as the reference and identify $M_1=M$.
Denoting $r_{i} = |\mathbf{x}_{i}|$ and $\hat{\rho}=|\mathbf{x}|$, and defining $\theta_{i}$ by $\mathbf{x} \cdot \mathbf{x}_{i} = \hat{\rho}\, r_{i} \cos \theta_{i}$, we can write
\begin{equation}
H=1+\mu+\frac{M}{\hat \rho^{D-3}}+\(\sum_{i=2}^{N} \frac{M_{i}}{|\mathbf{x}-\mathbf{x}_{i}|^{D-3}}-\mu \)
\end{equation}
where $\mu$ is the monopole contribution induced by the other BHs
\begin{equation}
\mu = \lim_{|x| \rightarrow 0} \frac{1}{V_{S^d}}\int \d \Omega_d  \(\sum_{i=2}^{N} \frac{M_{i}}{|\mathbf{x}-\mathbf{x}_{i}|^{D-3}}\) \, .
\end{equation}
Denoting
\be
H_0=1+\mu+\frac{M}{\hat \rho^{D-3}} \, ,
\ee
then we have
\begin{equation}
	H = H_0\left[1 + g(\hat{\rho})\right], \quad
{\rm with }\quad
 g(\hat{\rho}) = \frac{1}{\hat \rho^{D-3}H_0}\sum_{i=2}^{N} \sum_{j=1}^{\infty}m_{ij}(\theta_{i})\left(\frac{\hat{\rho}}{r_{i}}\right)^{D-3+j}\,,
\end{equation}
where $m_{ij}(\theta_{i}) = M_{i}C_{j}^{\left(\frac{D-3}{2}\right)}\left(\cos \theta_{i}\right)$ and $C^{(\alpha)}_{j}$ are Gegenbauer polynomials. The monopole $j=0$ is removed due to the introduction of $\mu$.

In the vicinity of the horizon of the first BH $\hat \rho \sim 0$ so we have at leading order
\begin{equation}
	H \approx \frac{M}{\hat{\rho}^{D-3}}\left[1 + g(\hat{\rho})\right], \quad g(\hat{\rho}) \approx \frac{1}{M}\sum_{i=2}^{N} \sum_{j=1}^{\infty}m_{ij}(\theta_{i})\left(\frac{\hat{\rho}}{r_{i}}\right)^{D-3+j} \, .
\end{equation}
Identifying
\be
\hat \rho^{D-3} = \frac{(D-3)}{(1+\mu)} r_H^{D-4} \rho \, ,
\ee
then the near-horizon metric to first order in perturbations takes the perturbed $\mathrm{AdS}_2 \times S^d$ form
\be
\d s^2 =  - \(1-2 g  \)  \frac{(D-3)^2}{(1+\mu)^2} \frac{\rho^2}{ r_H^2} \d t^2 +\(1+\frac{2 g }{(D-3)} \)   \frac{r_H^2}{(D-3)^2 \rho^2 } \d \rho^2 + r_H^2 \(1+\frac{2 g }{(D-3)} \) \d \Omega_d^2\,,\notag
\ee
with
\be
 g(\hat{\rho}) \approx \frac{1}{M}\sum_{i=2}^{N} \sum_{j=1}^{\infty}m_{ij}(\theta_{i}) \(\frac{(D-3)}{(1+\mu)} r_H^{D-4} \)^{\frac{(D-3+j)}{D-3}}\left(\frac{{\rho^{1+\frac{j}{(D-3)}}}}{r_{i}^{D-3+j}}\right) \, .
\ee
We see that orthonormal perturbations to the background metric scale as
\begin{equation}
	h \sim {\rho}^{\gamma} \, ,
\end{equation}
with
\begin{equation}
	\gamma = 1 + \frac{j}{D-3} \, ,\quad j \in \mathbb{N}.
\end{equation}
Identifying $j=\ell$, we see that these are the EM scalar modes $\gamma_{S+}$ in \eqref{eq: xrn critical exponents} (albeit written in a slightly different gauge). The monopole $\ell=0$ is absent, but the dipole $\ell=1$ is physical. By contrast for the gravity scalar modes, $\gamma_{S-}$, $\ell=1$ is pure gauge and $\ell=2$ is the first physical mode.

We see that in $D>4$ $\gamma$ is not generally an integer, which is merely a manifestation of the fact that horizons of multi-BH solutions in $D=4$ and $D>4$ are smooth and generically non-smooth  respectively \cite{Welch:1995dh, Candlish:2007fh, Gowdigere:2014aca, Gowdigere:2014cqa, Kimura:2014uaa}.

\section{Critical Exponents for Extremal Reissner-Nordstr\"om \label{app: xrn critical exponents}}

In this appendix, we collect the critical exponents of the near-horizon scaling solutions \eqref{eq: extremal solution} in GR and its IR and UV modifications considered in the main text.
We only state the more regular power $\gamma_{+}$ in \eqref{eq: hks solution}, as the other branch of the solutions can be switched off by finite boundary conditions at the horizon in the sub-extremal case.

\subsection{Pure GR \label{app: xrn critical exponents gr}}

Let us first consider pure GR.
We can use the effective potentials in \cite{Kodama:2003kk} to find the scaling for the tensor,
\ba
	\label{eq: xrn critical exponents}
		\gamma_{T} = -\frac{1}{2} + \frac{1}{2}\sqrt{1 + \frac{4\ell(\ell+D-3)}{(D-3)^{2}\sigma}}\,,
\ea
for the vectors,
\ba
		\gamma_{V\pm} = -\frac{1}{2} + \frac{1}{2} \sqrt{5 + \frac{4(\ell+1)(\ell+D-4)}{(D-3)^{2}\sigma} \pm 4 \sqrt{1 + \frac{2(\ell+1)(\ell+D-4)[1+(D-3)\sigma]}{(D-3)^{3}\sigma^{2}}}},\ \ \
\ea
and for the scalars,
\ba
		\gamma_{S\pm} = -\frac{1}{2} + \frac{1}{2}\sqrt{5 + \frac{4\ell(\ell+D-3)}{(D-3)^{2}\sigma} \pm 4 \sqrt{1 + \frac{4\ell(\ell+D-3)[1+(D-3)\sigma)]}{(D-3)^{2}(D-2)\sigma^{2}}}}.\label{eq: xrn critical exponentsSpm}
\ea
Note that
\begin{equation}
	\gamma_{M} > 0,\quad M\in \{T,V+,V-,S+\} \, ,
\end{equation}
for all $\sigma \in [0,\infty)$ and $\ell$.

This is however not the case for the gravitational scalar modes $\gamma_{S-}$.
In $D=4$ we find that $\gamma_{S-} \leq 2$ for
\begin{equation}
	\sigma \geq \frac{\ell(\ell+1)}{4}\left(1 - \sqrt{\frac{\ell^{2}+\ell+4}{3\ell(\ell+1)}}\right).
\end{equation}
For $\ell \geq 3$, the right-hand side of the inequality takes values in $[1,\infty)$ so this is never achieved for $\Lambda > 0$.
In $D \geq 5$ this is worse, and we even find $\gamma_{S-} \leq 0$ if
\begin{equation}
	\sigma \geq \frac{(D-2)k_{S}^{2}-4(D-3)^{2}}{2(D-4)(D-3)^{2}}\,,
\end{equation}
for $\ell > D-3$ in $\sigma \in [0,\infty)$.
For modes with $\ell \leq D-3$ this also holds for all $\sigma \in [1,\infty)$ and some values of $\sigma$ when $\Lambda >0$.
This is illustrated for $D=11$ in Figure~\ref{fig: scaling fixed dim}.

\begin{figure}
	\centering
	\includegraphics[scale=0.6]{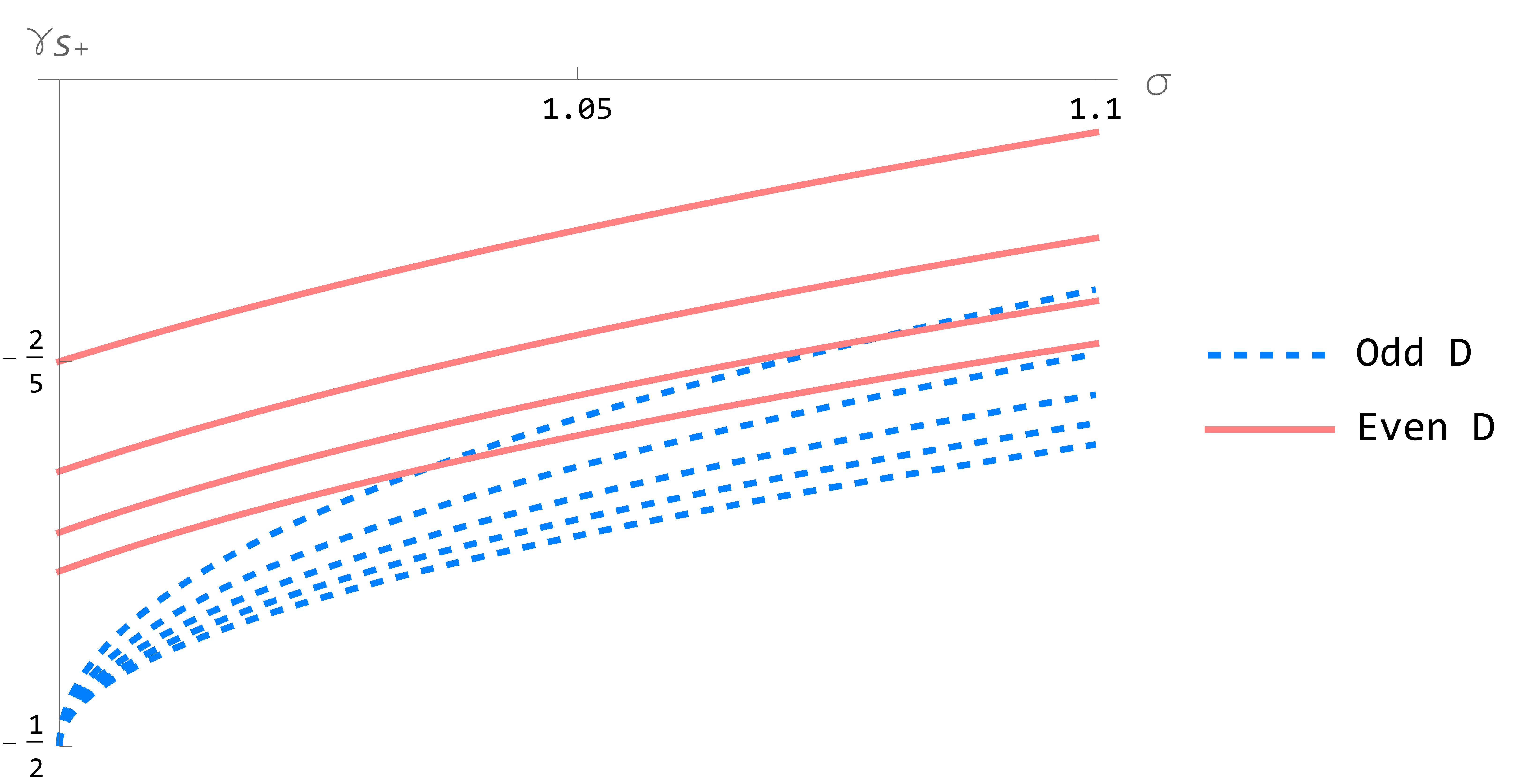}
	\caption{The (more regular) scaling dimension $\gamma_{+}$ for the most singular mode with $\ell=\lfloor(D-3)/2\rceil$ in various dimensions $D$. For odd $D$, the minimum $\gamma=-1/2$ is attained at $\sigma=1$. \label{fig: most singular scaling}}
\end{figure}

Further, note that $\gamma_{S-} \geq -1/2$ and this is saturated for $\ell=(D-3)/2$ at $\sigma = 1$.
As is shown in figure \ref{fig: most singular scaling}, this is possible only in odd $D$, although in even $D$,
$\ell=\lfloor(D-3)/2\rceil$ will achieve $\gamma = -\frac{1}{2} + \frac{1}{2(D-3)}$.
An exception to this is $D=5$, where the minimum is not attained since $\ell=1$ is not physical for gravitational scalar modes.

\subsection{IR: EFT Correction \label{subapp: xrn critical exponents eft}}

In this section, we will derive the critical exponents of gravitational scalar perturbations in Einstein-Maxwell theory with EFT corrections.
We shall focus on the leading four-derivative corrections given by the action
\begin{equation}	
	\begin{aligned}
		\kappa S_{\text{EFT}} &=\frac{1}{\Lambda^2} \int \dd^{5}x \sqrt{-g} \bigg[c_{1} R^{2} + c_{2} R_{\mu\nu}R^{\mu\nu} + c_{3} R_{\mu\nu\rho\sigma}R^{\mu\nu\rho\sigma}+c_{4}R \widehat F^{2} \\
		&\quad + c_{5} R_{\mu\nu} \widehat F\indices{^{\mu}_{\lambda}} \widehat F^{\nu\lambda} + c_{6}R_{\mu\nu\rho\sigma} \widehat F^{\mu\nu} \widehat F^{\rho\sigma} + c_{7} (\widehat F^{2})^{2} + c_{8} \widehat F\indices{_{\mu}^{\nu}} \widehat F\indices{_{\nu}^{\rho}} \widehat F\indices{_{\rho}^{\sigma}} \widehat F\indices{_{\sigma}^{\mu}}\bigg].
	\end{aligned}
\end{equation}
To find the critical scaling exponents, we follow the procedure in appendix \ref{app: nh decomposition} and start directly within the extremal near-horizon geometry.
For simplicity, we restrict ourselves to $D=5$.

In particular, the Ansatz \eqref{eq: freund-rubin ansatz} satisfies the background equations of motion for
\begin{eqnarray}	
\ANH &=& \frac{2\kappa\FNH^{2}}{1+2\sigma} \bigg\{\sigma  + \frac{1}{r_{H}^{2}\Lambda^{2}}\frac{1}{2(1+2\sigma)}\big[2(-3+4\sigma)(-1+4\sigma+8\sigma^{2})c_{1}+2(1-4\sigma+16\sigma^{3})c_{2} \notag \\
			&+&2(1-4\sigma+32\sigma^{3})c_{3}+4(1-4\sigma+24\sigma^{3})c_{4}+4\sigma(1+2\sigma)(-1+6\sigma)(c_{5}+2c_{6}) \notag\\
			&+&4(1+2\sigma)^{2}(-1+4\sigma)(2c_{7}+c_{8})\big]\bigg\}	\label{eq: dim-6 background} \\
\BNH &=& \frac{4}{r_{H}^{2}}\bigg\{\sigma+\frac{1}{r_{H}^{2}\Lambda^{2}}\big[2(3-4\sigma)c_{1}+2(c_{2}+c_{3})
+4(1-4\sigma^{2})c_{4}-4\sigma(1+2\sigma)(c_{5}+2c_{6}) \notag \\
			&-&4(1+2\sigma)^{2}(2c_{7}+c_{8})\big]\bigg\}.
\end{eqnarray}
Using the decomposition in appendix \ref{subapp: nh perturbations}, we find that perturbations on top of this background take the form of the scaling solutions \eqref{eq: critical scaling} with critical exponents in the form of \eqref{eq: critical exponent resummed}.
More specifically, we find that
\begin{equation}
	\geft = \frac{1}{\Lambda^{2}r_{H}^{2}} \Delta \sum_{i=1}^{8} c_{i} \delta \gamma_{i},
\end{equation}
where
\begin{subequations}
	\label{eq: xrn critical exponent dim-6}
	\begin{align}
		\delta \gamma_{1} &= -576(z-2)z(4\sigma-3)\sigma\\
  &-16k_{S}^{2}\left[(-8\sigma^{2}-40\sigma+27)z-20\sigma+5\right]
  -16k_{S}^{4}(1+2\sigma)\notag\\
			\delta \gamma_{2} &= 576 (z-2)z(1+2\sigma+4\sigma^{2}) \sigma \\
   &-16k_{S}^{2}\left[(56\sigma^{2}+28\sigma+9)z+48\sigma^{2}+80\sigma+31\right]
			+40k_{S}^{4}(1+2\sigma)\notag  \\
			\delta \gamma_{3} &= 576(z-2)z(1+4\sigma+16\sigma^{4})\sigma \\
   &+16k_{S}^{2}\left[-(256\sigma^{2}+20\sigma+9)z-144\sigma^{2}-280\sigma-47\right]
			+64 k_{S}^{4}(-1+7\sigma)\notag \\
		\delta \gamma_{4} &=-1152 (z-2)z(-1+4\sigma^{2})\sigma\\
  &+96k_{S}^{2}(1+2\sigma)\left[(10\sigma-3)z+2\sigma+3\right]-96k_{S}^{4}(1+2\sigma) \notag\\
\delta \gamma_{5} &= 576(z-2)z \sigma^{2} \left[(z-6)\sigma-2\right]\\
&-48k_{S}^{2}\left[9z^{2}\sigma^{2}-12z \sigma(1+4\sigma)-4(\sigma^{2}+6\sigma+2)\right]  -36k_{S}^{4}\left[(-3z+8)\sigma+2\right]-9k_{S}^{6} \notag \\
			\delta \gamma_{6} &= 2304 (z-2)z \left[(9z-14)+2\right]\sigma^{2}\\
   &-192k_{S}^{2}\left[(9z^{2}-32z-7)\sigma^{2}-(z+19)\sigma-2\right]
+48k_{S}^{4}\left[(9z-14)\sigma+2\right]-36k_{S}^{6}\notag \\
			\delta \gamma_{7} &= -2304(z-2)z(1+2\sigma^{2})\sigma+192k_{S}^{2}(1+2\sigma)\left[3z(1+6\sigma)+2\sigma+13\right] \\
			&-576k_{S}^{4}(1+2\sigma)
			=  2\delta \gamma_{8},
	\end{align}
\end{subequations}
and
\begin{equation}
	\begin{aligned}
		\Delta &= \frac{1}{3} (\ggr+1)(\ggr+2) \big\{-48(1+\ggr)^{2}(2+\ggr)^{2}(-1+2\ggr)\sigma^{2} \\
		&+8k_{S}^{2}\left[(6\ggr^{2}+10\ggr+3)\sigma-4\ggr-6\right]+3k_{S}^{4}(2\ggr+3)\big\}^{-1}.
	\end{aligned}
\end{equation}
For convenience we have also further defined
\begin{equation}
	z = \ggr^{2} + \ggr  = 1 + \frac{k_{S}^{2}}{4\sigma} - \sqrt{1 + \frac{1+2\sigma}{3\sigma^{2}}k_{S}^{2}}
\end{equation}
The asymptotically flat limit is given by sending $\sigma \rightarrow 1$ and reproduces the expressions in \cite{Horowitz:2023xyl}.

\subsection{UV: Einstein-Maxwell-Dilaton \label{subapp: xrn critical exponents uv}}

In this section, we derive the critical exponents in Einstein-Maxwell-Dilaton theory with action \eqref{eq: partial uv completion tower}, treating everything perturbatively in $\alpha$.
We are only interested in the critical exponents, so we follow the procedure in appendix \ref{app: nh decomposition} and start directly within the extremal near-horizon geometry.

For simplicity, we restrict ourselves to $D=5$.
Then, the Ansatz \eqref{eq: freund-rubin ansatz} satisfies the background equations of motion for
\begin{subequations}	
	\label{eq: dilaton background}
	\begin{align}
		\phi_{i} &= \frac{\alpha_i \BNH \kappa \FNH^{2}}{2m_{i}^{2}}, \\
		\ANH &= \frac{\kappa \FNH^{2}}{2}\left[\frac{4}{3} \frac{1+3\varsigma}{1+2\varsigma} + \frac{3}{r_{H}^{2}\meff^{2}}\left(1+4\varsigma\right)\right], \\
  \BNH &= \frac{4}{r_{H}^{2}} \left(1+3\varsigma\right) - \frac{9}{\meff^{2}r_{H}^{4}}\left(1+2\varsigma\right)^{2}\,,
	\end{align}
\end{subequations}
where $\meff$ is defined in \eqref{effectivemass}. Since each scalar is only non-zero at order $\mathcal{O}(\alpha)$, the corrections in the perturbation equations will come in at $\mathcal{O}(\alpha^{2})$.
Thus the first non-trivial corrections to the critical exponents will come in at $\mathcal{O}(\alpha^{2})$.

In the end, we are only interested in the gravitational scalar modes, as these are the ones that exhibit the singular behaviour we are interested in.
The decomposition in appendix \ref{subapp: nh perturbations} then renders the perturbation equations algebraic in the constants, and the critical exponent takes the form \eqref{eq: dilaton critical exponent form} with $\ggr$ as given in \eqref{eq: xrn critical exponents} and
\begin{equation}
	\label{eq: xrn critical exponent dilaton}
	\frac{1}{\Delta}\delta \gamma = -9k_{S}^{2}(1+2\sigma)(-4z\sigma-4+k_{S}^{2})\chi+\frac{6}{\meff^{2}r_{H}^{2}}(1+2\sigma)^{2}\left[-12(z-2)z\sigma+k_{S}^{2}(1+3z)\right],
\end{equation}
where
\begin{equation}
	\chi = \sum_{i=1}^{N} \frac{2 \alpha_i^2}{r_{H}^{2}m_{i}^{2}+k_{S}^{2}-4z\sigma}.
\end{equation}

\section{Einstein-Maxwell Action Expansion \label{app: em action expansion}}

An expansion of the Einstein-Maxwell action \eqref{eq: em action} in the metric and field strength perturbations $h$ and $\delta F$ takes the form
\begin{equation}	
	S_{\text{EM}}[g,F] = \sum_{N=0}^{\infty}S^{(N)}[\bar{g},h;\bar{F},\delta F]
	\label{eq: em action expansion}
\end{equation}
where $S^{(N)}$ has $N$ factors of the metric or field strength perturbations in total.
When evaluated on the background, $S^{(0)}$ is just some constant and $S^{(1)}$ is a boundary contribution at most.
Furthermore, when evaluated on-shell of perturbations, the quadratic action $S^{(2)}$ is also only a boundary contribution.

The explicit form of the cubic  action is
\begin{eqnarray}
\label{eq: em action cubic}
\frac{1}{3!\kappa^{1/2}}S^{(3)}[h,\delta F]
  &=& \int \dd^{D}x \sqrt{-g} \,\bigg\{ h^{\mu\nu} \bigg[h^{\rho\sigma}\nabla_{\sigma}\nabla_{[\nu}h_{\rho]\mu}
+h\indices{_{\mu}^{\rho}}\left(\nabla^{\sigma}\nabla_{[\rho}h_{\sigma]\nu}
+\nabla_{\rho}\nabla_{[\sigma}h\indices{_{\nu]}^{\sigma}}\right) \qquad \\
  &+&\nabla_{\nu} h\indices{_{\mu}^{\rho}}\nabla_{[\sigma}h\indices{_{\rho]}^{\sigma}}
+ \nabla_{[\nu} h_{\rho]\mu}\nabla_{\sigma}h^{\rho\sigma}+\nabla_{\nu}h\indices{^{\sigma}_{[\rho}}\nabla_{\sigma]}h\indices{_{\mu}^{\rho}}+\frac{1}{2}h\nabla_{\rho}h\indices{_{\mu}^{\rho}}\nabla_{\sigma}h\indices{_{\nu}^{\sigma}} \notag \\
&-&\frac{1}{2}\nabla_{\rho}h\indices{^{\sigma}_{[\nu}}\nabla^{\rho}h_{\sigma]\mu}-\frac{1}{2}\nabla^{\rho}h\indices{^{\sigma}_{\nu}}\nabla_{[\rho}h_{\sigma]\mu}
		-\frac{1}{4}\nabla_{\sigma}h_{\nu\rho}\nabla^{\sigma}h\indices{_{\mu}^{\rho}} -\frac{3}{8}\nabla_{\mu}h^{\rho\sigma}\nabla_{\nu}h_{\rho\sigma}\notag \\
&+&\frac{1}{8}\nabla_{\mu}h\nabla_{\nu}h
- \frac{1}{2} h\indices{_{\mu}^{\rho}}h\indices{_{\nu}^{\sigma}}\bar{R}_{\rho\sigma} + \frac{1}{2} \delta F\indices{^{\rho}_{\nu}}\delta F_{\rho\mu}\bigg] \notag \\
		&-&\kappa^{1/2} \delta F^{\mu\nu}\bigg[h^{\rho\sigma}\bar{F}_{\mu\rho}h_{\nu\sigma}+\frac{1}{2}\bar{F}^{\rho\sigma}h_{\mu\rho}h_{\nu\sigma}\bigg] + \kappa\bar{F}^{\mu\nu}h_{\sigma\lambda}h\indices{_{\nu}^{\lambda}}\bar{F}\indices{^{\rho}_{(\mu}}h_{\sigma)\rho}  \bigg\}\notag
\end{eqnarray}
and that at quartic order,
\begin{equation}
	\label{eq: em action quartic}
	\begin{aligned}
		\frac{1}{4!\kappa }&S^{(4)}[h,\delta F] = \int d^{D}x \sqrt{-g}\, \bigg\{ h^{\mu\nu}h\indices{_{\mu}^{\rho}}h\indices{_{\nu}^{\sigma}}\left(-\nabla^{\lambda}\nabla_{[\sigma}h_{\lambda]\rho}-\nabla_{\sigma}\nabla_{[\lambda}h\indices{_{\rho]}^{\lambda}} +\frac{1}{2}h\indices{_{\rho}^{\lambda}}\bar{R}_{\sigma\lambda}+\frac{1}{2}h^{\kappa\lambda}\bar{R}_{\rho\kappa\sigma\lambda}\right) \\
		&+h^{\mu\nu}h\indices{_{\mu}^{\rho}}h^{\sigma\lambda}\left(-\nabla_{\lambda}\nabla_{[\rho}h_{\sigma]\nu}-\nabla_{\rho}\nabla_{[\lambda}h_{\nu]\sigma}+\frac{1}{2}h\indices{_{\sigma}^{\kappa}}\bar{R}_{\nu\rho\kappa\lambda}+\frac{1}{2}h\indices{_{\nu}^{\kappa}}\bar{R}_{\rho\lambda\sigma\kappa}\right) \hspace{-2cm}\\	&+h^{\mu\nu}h\indices{_{\mu}^{\rho}}\bigg(-2\nabla^{\lambda}h_{\sigma[\lambda}\nabla_{\rho]}h\indices{_{\nu}^{\sigma}}+\nabla_{(\rho}h\nabla_{\sigma)}h\indices{_{\nu}^{\sigma}}-\nabla^{\sigma}h\indices{_{\nu}_{[\sigma}}\nabla^{\lambda}h_{\rho]\lambda}-\nabla_{\rho}h^{\sigma\lambda}\nabla_{[\lambda}h_{\nu]\sigma}\hspace{-2cm} \\
		&-\frac{1}{2}\nabla_{(\lambda}h_{\sigma)\rho}\nabla^{\lambda}h\indices{_{\nu}^{\sigma}}-\frac{1}{4}\nabla_{\nu}h\indices{^{(\sigma}_{\lambda}}\nabla_{\rho}h\indices{^{\lambda)}_{\sigma}}-\frac{1}{4}\nabla_{\sigma}h\nabla^{\sigma}h_{\nu\rho}\bigg) \\
		&+h^{\mu\nu}h^{\rho\sigma}\bigg(-2\nabla_{\nu}h_{\mu[\sigma}\nabla_{\lambda]}h\indices{_{\rho}^{\lambda}}+\nabla_{\nu}h_{\mu(\lambda|}\nabla_{\sigma}h\indices{_{|\rho)}^{\lambda}}+\nabla_{(\sigma|}h\indices{_{\rho}^{\lambda}}\nabla_{|\lambda)}h_{\mu\nu} -\nabla_{\sigma}h\indices{_{\nu}^{\lambda}}\nabla_{[\lambda}h_{\rho]\mu} \hspace{-2cm} \\
		&+\frac{1}{2}\nabla_{[\sigma}h_{\nu]\lambda}\nabla_{\rho}h\indices{_{\mu}^{\lambda}}-\frac{1}{4}\nabla_{\lambda}h_{\rho[\sigma}\nabla^{\lambda}h_{\mu]\nu}+\frac{1}{4}\nabla_{\lambda}h_{\nu\sigma}\nabla^{\lambda}h_{\mu\rho}-\frac{1}{4}\nabla_{\rho}h_{\mu\nu}\nabla_{\sigma}h\bigg) \\
		&-\frac{1}{2}h^{\mu\nu}h_{\rho\mu}\delta{F}\indices{^{\rho}_{\sigma}}\delta{F}\indices{_{\nu}^{\sigma}}-h^{\mu\nu}h_{\rho\sigma}\delta{F}\indices{_{\mu}^{\rho}}\delta{F}\indices{_{\nu}^{\sigma}} \\
  &-\kappa^{1/2} \delta F^{\mu\nu}\bigg[\bar{F}^{\rho\sigma}h\indices{_{\sigma}^{\lambda}}h_{\mu[\lambda}h_{\rho]\lambda}+\bar{F}\indices{^{\rho}_{\mu}}h\indices{_{\rho}^{\lambda}}h_{\sigma\lambda}h\indices{_{\nu}^{\sigma}}\bigg] \\
		&-\frac{1}{2}\kappa\bar{F}^{\mu\nu}\bigg[h\indices{_{\sigma}^{\lambda}}h_{\kappa\lambda}\left(\bar{F}^{\rho\sigma}h_{\mu\rho}h\indices{_{\nu}^{\kappa}}+\bar{F}\indices{_{\mu}^{\rho}}h\indices{_{\nu}^{\sigma}}h\indices{_{\rho}^{\kappa}}\right) + \frac{1}{2}\bar{F}^{\rho\sigma}h\indices{_{\mu}^{\kappa}}h\indices{_{\nu}^{\lambda}}h_{\rho\kappa}h_{\sigma\lambda}\bigg]\bigg\}
	\end{aligned}
\end{equation}
with $\bar{R}$ being the Riemann tensor evaluated on the background solution.

\bibliographystyle{JHEP}
\bibliography{Bibliography.bib}

\end{document}